\documentclass[aps,showpacs,preprint,preprintnumbers,amsmath,amssymb,nofootinbib]{revtex4}
\usepackage{epsfig}
\usepackage{graphicx}
\usepackage{dcolumn}
\usepackage{bm}
\usepackage{feynmp}
\usepackage{slashed}

\def\beq{\begin{equation}}
\def\eeq{\end{equation}}

\newcommand{\bea}{\begin{eqnarray}}
\newcommand{\eea}{\end{eqnarray}}

\def\eeqn{\end{equation}}
\newcommand\iden{\leavevmode\hbox{\small1\normalsize\kern-.33em1}}

\let\jnfont=\rm
\def\NPB#1,{{\jnfont Nucl.\ Phys.\ B }{\bf #1},}
\def\PLB#1,{{\jnfont Phys.\ Lett.\ B }{\bf #1},}
\def\EPJC#1,{{\jnfont Eur.\ Phys.\ Jour.\ C }{\bf #1},}
\def\PRD#1,{{\jnfont Phys.\ Rev.\ D }{\bf #1},}
\def\PRL#1,{{\jnfont Phys.\ Rev.\ Lett.\ }{\bf #1},}
\def\MPLA#1,{{\jnfont Mod.\ Phys.\ Lett.\ A }{\bf #1},}
\def\JPG#1,{{\jnfont J.\ Phys.\ G }{\bf #1},}
\def\CTP#1,{{\jnfont Commun.\ Theor.\ Phys.\ }{\bf #1},}
\def\JHEP#1,{{\jnfont JHEP \ }{\bf #1},}
\def\NPPS#1,{{\jnfont Nucl.\ Phys.\ Proc.\ Suppl.\ }{\bf #1},}

\begin{document}


\title{Phenomenological Aspects of R-parity Violating Supersymmetry
with A Vector-like Extra Generation}

\author{{Xue Chang}, {Chun Liu}, {Yi-Lei Tang}}
\affiliation{
State Key Laboratory of Theoretical Physics,
Institute of Theoretical Physics, Chinese Academy of Sciences,
P.O. Box 2735, Beijing 100190, China}
\email{chxue@itp.ac.cn, liuc@mail.itp.ac.cn, tangyilei@itp.ac.cn}

\begin{abstract}
Phenomenological analysis to the R-parity violating supersymmetry with
a vector-like extra generation is performed in detail.  It is found
that, via the trilinear couplings, the correct neutrino spectrum can
be obtained.  The Higgs mass rises to 125 GeV by new up-type Yukawa
couplings of vector-like quarks with no need of very heavy
superpartners.  Phenomena of new heavy fermions at LHC are predicted.
\end{abstract}


\maketitle

\section{Introduction}

Recently a standard model (SM) Higgs-like particle with a mass of
$125-126$ GeV was discovered \cite{atlas-cms}.  In the paradigm of the
weak scale supersymmetry (SUSY) which aims at the naturalness of the
electro-weak scale, however, such a Higgs mass brings in tensions,
especially the minimal SUSY SM (MSSM).  Nonminimal and still natural
scenarios of SUSY are thus motivated.  One of them is the MSSM with a
vector-like generation \cite{ns-vc,nath,liu,Higgs2}.  It gives the
right Higgs mass naturally, is consistent with precision electroweak
measurements, and has a rich phenomenology \cite{ns-vc,nath,liu,Higgs2,vc}.
In the framework of SUSY, vector-like fermions can also be motivated
by other theories beyond SM, such as SUSY extension with extra-dimensions
or with composite states \cite{extra-d}. So it is worth asking the question
whether such a scenario also provides explanations to other problems
such as neutrino masses.

Neutrino oscillations are the undoubted new physics beyond the SM.  Daya
Bay \cite{daya} and RENO \cite{reno} experiments recently discovered a
relatively large $\theta_{13}\simeq 8.8^\circ\pm 0.8^\circ$.  Within the
framework of SUSY, in the absence of R-parity conservation, neutrino
masses and mixings can be generated from lepton number violating (LPV)
couplings \cite{rpv1}.  This approach was extensively studied before
\cite{neutrino-rpv}.  It is known that all the neutrino experimental
results, including that of oscillation phenomena like the large
atmospheric mixing angle $\theta_{23}$, the hierarchy of oscillation
frequencies  $\Delta m^2_{21}\ll \Delta m^2_{32}$ and the smallness of
$\theta_{13}$, can be understood in three generation LPV MSSM.  However,
this needs some special requirements for relevant coupling constants and
mass parameters.

Combining both considerations above, we will work in the LPV MSSM with a
vector-like extra generation \cite{liu}.  While this model takes the
vector-like slepton doublets as the two Higgs doublets needed for the
electroweak symmetry breaking, the SM-like Higgs mass can be naturaly
125 GeV \cite{Higgs2}.  Extra trilinear LPV couplings between ordinary
fermions and vector-like fermions provide a much larger parameter space to explain
neutrino pheomena right.

In this paper, phenomenological aspects of the model will be analyzed.
In Sect. II, we make a brief review of the model.
In Sect. III, neutrino masses are calculated.  For the neutrino physics,
noting the enlarged parameter space, we consider trilinear LPV couplings
carefully.  One-loop contribution to neutrino masses due to new
trilinear LPV couplings is calculated, theoretical analysis are
performed and numerical results are shown in detail.  Besides, we
analyze the SM-like Higgs mass and explicitly show that it can be
increased to 125 GeV by two new Yukawa couplings of the up-type Higgs
with vector-like quarks in Sect. IV.  The LHC phenomenology of the new
fermions is analyzed in Sect. VI.  The summary and discussions are given
in the last section.

\section{A BRIEF REVIEW OF THE MODEL}

This model\cite{liu} is SUSY and SM gauge invariant, and R-parity violation
with baryon number conservation
is assumed.  For the matter content, in addition to the ordinary 3
generations (3G), a vector-like generation is introduced in.  Without
R-parity conservation, this can be also
thought as that there are $4+1$ chiral generations, where '4' stands
for four chiral generations with SM quantum numbers and '1' for another
chiral generation with opposite quantum numbers.
The 4 chiral generations with same quantum numbers mix.  The '1' has
Dirac masses with only one combination of the '4', thus, there are
always SM required three massless chiral generations and one massive
vector-like generation.

In terms of mass eigenstates (before
electroweak symmetry breaking), the massive sleptons in the vector-like
generation are taken as the two Higgs doublets.  New particles beyond the
MSSM are the following with quantum numbers under
SU(3)$_c\times$SU(2)$_L\times$U(1)$_Y$,
\begin{equation}
\begin{array}{c}
E_4^c(1, 1,2)~,~E_H^c(1,1,-2)~,~ Q_4(3, 2,~ \frac{1}{3})~,~
Q_H(\bar{3}, 2, -\frac{1}{3})~, \\[3mm]
U_4^c(\bar{3}, 1, -\frac{4}{3})~,~ U_H^c(3, 1, \frac{4}{3})~,~
D_4^c(\bar{3}, 1, \frac{2}{3})~,~ D_H^c(3, 1,-\frac{2}{3})~.
\end{array}
\nonumber
\end{equation}
The superpotential is conveniently written as
\begin{equation}
{\mathcal W} = {\mathcal W}_0 + {\mathcal W}_{{\not L}} \,,
\end{equation}
where ${\mathcal W}_0$ and ${\mathcal W}_{{\not L}}$ stand for that
with lepton number conservation and LPV, respectively,
\begin{equation}
\begin{array}{lll}
{\mathcal W_0} & = & \mu H_u H_d + \mu^e E_4^c E_H^c
 + \mu^Q Q_4 Q_H + \mu^U U_4^c U_H^c
 + \mu^D D_4^c D_H^c
 + y_{ij}^l L_i H_d E_j^c
 + y_{ij}^d Q_i H_d D_j^c\nonumber \\
&&  + y_{ij}^u Q_i H_u U_j^c + y_i^E L_i H_d E_4^c
 + y_i^{Q\prime} Q_4 H_d D_i^c + y_i^D Q_i H_d D_4^c
 + y^{QD} Q_4 H_d D_4^c
 + y_i^U Q_i H_u U_4^c\nonumber \\
&& + y_i^Q Q_4 H_u U_i^c + y^{QU} Q_4 H_u U_4^c + y^H Q_H H_d U_H^c
 + y^{H'} Q_H H_u D_H^c \,,
\end{array}
\end{equation}
and
\begin{eqnarray}
{\mathcal W}_{{\not L}}\
             \supset && \lambda_{ijk} L_i L_j E_k^c
                + \lambda'_{ijk} Q_i L_j D_k^c
                + \lambda_{ij}^E L_i L_j E_4^c
                + \lambda_{ij}^Q Q_4 L_i D_j^c \nonumber\\
             &&  + \lambda_{ij}^D Q_iL_jD_4^c
                + \lambda_i^{QD} Q_4 L_i D_4^c
                + \lambda^H_i Q_H L_i U_H^c \,.
\end{eqnarray}
where $L_i$, $Q_i$, $E_i^c$, $D_i^c$, $U_i^c$, i=1-3, are the first three
generation $SU(2)_L$ doublet leptons, doublet quarks, singlet charged leptons,
singlet down-type quarks and singlet up-type quarks, respectively.
$H_u$ and $H_d$ are the up-type and down-type
Higgs. Note that the term $Q_HH_uD_H^c$ in ${\mathcal W_0}$ was missed in Ref.
\cite{liu}.\footnote{It modifies the down-type fermion mass matrix and
scalar mass-squared matrix.  Correct ones, as well as the resulting
mixing matrix are given in the Appendix A.}
And in ${\mathcal W}_{{\not L}}$ interactions of purely
singlets are omitted, which are irrelevant to our study.

By assuming universality of the mass-squared terms, the alignment
of the $B$ terms the soft mass terms and the trilinear soft terms
of all fermion's superpartners in the model are
\begin{eqnarray}
-{\mathcal L}&\supset& M^2\tilde{L}_i^{\dag}\tilde{L}_i+M^2H_d^{\dag}H_d
            + M_h^2H_u^{\dag}H_u + M_E^2\tilde{E}_m^{c\dag}\tilde{E}_m^c
                      + M_Q^2\tilde{Q}_m^{\dag}\tilde{Q}_m
                      + M_U^2\tilde{U}_m^{c\dag}\tilde{U}_m^c\nonumber\\
            &        &+ M_D^2\tilde{D}_m^{c\dag}\tilde{D}_m^c
                      + M_{EH}^2\tilde{E}_H^{c*}\tilde{E}_H^c
                      + M_{QH}^2\tilde{Q}_H^{\dag}\tilde{Q}_H
                      + M_{UH}^2\tilde{U}_H^{c*}\tilde{U}_H^c
                      + M_{DH}^2\tilde{D}_H^{c*}\tilde{D}_H^c\\
            &        &+(B\mu H_dH_u+B^e\mu^e\tilde{E}_4^c\tilde{E}_H^c
                      +B^Q\mu^Q\tilde{Q}_4\tilde{Q}_H
                      +B^U\mu^U\tilde{U}_4^c\tilde{U}_H^c
                      +B^D\mu^D\tilde{D}_4^c\tilde{D}_H^c + \mathrm{h.c.})\,.\nonumber
\end{eqnarray}

Proper values of the new $B^{Q,U,D}\mu^{Q,U,D}$ terms are set to
avoid unwanted color symmetry and purely $U(1)_Y$ symmetry breaking,
see Eq. (11, 12) in paper \cite{liu}, therefore EWSB in our model is
just the same as in MSSM. After EWSB, the specific fermion mass matrixes
and sfermion mass-squared matrixes are given in Appendix A.

\section{NEUTRINO MASSES AND MIXINGS}

LPV results in nonvanishing neutrino masses.  In this model, in addition
to traditional R-parity violation in the MSSM, a lot more bilinear and
trilinear LPV interactions are brought in through the vector-like
generation.  In this work, the trilinear R-parity violating interactions
will be studied.  To avoid complication due to too many LPV sources,
sneutrino VEVs will not be considered.  There are several reasons for
this.First, we can phenomenologically assume the universality of the
soft SUSY breaking mass terms at the weak scale, to avoid dangerously
large flavor changing neutral currents (FCNCs), without considering any
UV completion of the model.  In that case, because of the alighnment in
bilinear terms of the superpotential and that of soft terms, R-parity
violating bilinear terms can be rotated away via field redefinition, and
sneutrino vacuum expectation values (VEVs) vanish in the physical basis.
The second reason is from consideration of underlying models.  SUSY
breaking is introduced effectively in our model, it can result from
gauge mediated SUSY breaking.  Then the messenger scale can be as low as
100 TeV, even if the universality scale is at the SUSY breaking
messenger scale, the running effect is small, and the bilinear LPV is
not important compared to the trilinear ones.
Finally, small sneutrino VEVs can be included in the analysis
nevertheless in future works, after the role of new trilinear
LPV interactions gets a thorough understanding.

The trilinear LPV Lagrangian relevant to neutrino masses is from
${\mathcal W}_{{\not L}}$,
\begin{eqnarray}
   \cal{L} & \subset &
   -\lambda_{ijk}(\tilde{l}_{kR}^* \bar{\nu}_{iR}^c l_{jL}
   +\tilde l_{jL}\bar l_{kR}\nu_{iL})
   - \lambda '_{ijk} (\tilde d^* _{kR}\bar \nu^c_{iR}d_{jL}
   + \tilde d_{jL}\bar d_{kR}\nu_{iL})\nonumber \\
   &&-\lambda_{ij}^E(\tilde{E}_4^c \bar{\nu}_{iR}^c l_{jL}
   +\tilde l_{jL} E_4^{cT} \nu_{iL})
   -\lambda_{ij}^Q(\tilde d^* _{kR}\bar \nu^c_{iR}Q_4
   +\tilde Q_4 \bar d_{kR}\nu_{iL}) \\
  && -\lambda_{ij}^D(\tilde D_4^c \bar \nu^c_{iR}d_{jL}
   +\tilde d_{jL}\bar D_4^{cT}\nu_{iL})
   -\lambda_i^{QD}(\tilde D_4^c \bar \nu^c_{iR}Q_4
   +\tilde Q_4 \bar D_4^{cT} \nu_{iL})\nonumber \\
   &&-\lambda^H_i(\tilde U_H^c \bar \nu^c_{iR} Q_H
   +\tilde Q_H \bar U_H^{cT} \nu_{iL})
   +\mathrm{h.c.}\nonumber\,.
\end{eqnarray}
where $\bar{\nu}_{iR}^c$ stands for the left-hand neutrino.

The 7 types of trilinear LPV interactions in the above equation
induce 14 types of one-loop diagrams contributing to the neutrino
spectrum, which are proportional to
$\lambda \lambda,~\lambda' \lambda',~\lambda^E \lambda^E$,
$\lambda\lambda^E,~ \lambda^Q \lambda^Q$, $\lambda^Q \lambda^D,
~\lambda^D \lambda^D$,
$\lambda^H \lambda^H,~\lambda^{QD} \lambda^{QD} ,~\lambda'
\lambda^Q,~\lambda' \lambda^D,~\lambda' \lambda^{QD},$
$\lambda^Q\lambda^{QD},~\lambda^D \lambda^{QD}$, respectively.
The Feynman diagrams and the corresponding analytical results
are shown in Fig. 1 in Appendix B. For simplicity and without losing our purpose,
in the Yukawa interactions of $\mathcal{W}_0$ we assume that only
$y^E$, $y^{Q'}$, $y^{Q'}$, $y^D$, $y^U$, $y^H$, $y^{H'}$ are
nonvanishing, that is vector-like particles have
Yukawa interactions only with the third generation.  Thus, the
vector-like generation has little constraints from the collider
phenomenology.

Before starting to analyze the neutrino mass spectrum, some
assumptions are introduced in order to control the parameter space
and get relatively simple analytical result. Since four new
up-type Higgs Yukawa couplings $y^U,~y^Q,~y^{QU},~y^{H'}$ and five
new down-type Higgs Yukawa couplings $y^E,~y^D,~y^{Q'},~y^{QD},~y^{H}$
appear in our model, and among which
$y^{QD},~y^{QU},~y^{H},~y^{H'}$ provide the mass mixings between
vector-like generations, and further more, they have infrared
quasi-fixed point \cite{Higgs2}, so we assume $y^{QD}=y^{H}=0$ and $y^{QU}\sim
y^{H'}\equiv y^t_V\leq1$. We also set  $y^D=y^{Q'}=0$, $y^E<0.04$ and $y^U\sim
y^Q\equiv y^t_{34}\leq0.08$. In other words, we neglect all new
down-type Higgs Yukawa couplings in quark sectors while consider
all of the new up-type Higgs Yukawa couplings only and take
$y^t_{34}\ll y^t_V$, which is a reasonable assumption.

Basing on the above assumptions,
contributions from $\lambda \lambda$, $\lambda' \lambda'$ type
diagrams can be simplified to the familiar forms
\cite{neutrino1,neutrino2,neutrino3}
\begin{eqnarray}
 &&M^{\nu}_{ij} |_{\lambda \lambda}\ \simeq\ \frac{1}{8 \pi^2}
  \lambda_{i33} \lambda_{j33}\,m_\tau \sin\alpha_{\tilde{\tau}}\cos\alpha_{\tilde{\tau}}
  \ln\frac{\tilde{\tau}_R}{\tilde{\tau}_L},\nonumber\\
 &&M^{\nu}_{ij} |_{\lambda' \lambda'}\ \simeq\ \frac{3}{8 \pi^2}(
  \lambda'_{i33} \lambda'_{j33} m_{b}
  \sin\alpha_{\tilde{b}}\cos\alpha_{\tilde{b}}\ln\frac{\tilde{b}_R}{\tilde{b}_L}+
  \lambda'_{i23} \lambda'_{j32} m_{s}
  \sin\alpha_{\tilde{b}}\cos\alpha_{\tilde{b}}\ln\frac{\tilde{b}_R}{\tilde{b}_L}\nonumber\\
  &&~~~~~~~~~~~~+\lambda'_{i32} \lambda'_{j23}\,m_{b}
  \sin\alpha_{\tilde{s}}\cos\alpha_{\tilde{s}}\ln\frac{\tilde{s}_R}{\tilde{s}_L}),
\end{eqnarray}
where in the first equation, we only keep the dominant
contributions and in the second equation, we keep the dominant and
subdominant ones. $\alpha_{\tau}$, $\alpha_{b}$, $\alpha_{s}$,
$\alpha_{t}$ are the angles of the corresponding 2$\times$2
$\tilde{\tau}_{L(R)}, \tilde{b}_{L(R)}, \tilde{s}_{L(R)},
\tilde{t}_{L(R)}$ unitary matrices. Unfortunately, the other
equations, (B3)-(B14) in Appendix B, can not be simplified by following
similar process because there are mixings between different
vector-like generations. So these can only be analyzed numerically
and will be discussed later.

At last, without loss of generality, among all 7 types of LPV
trilinear couplings, we take 4 of them, $\lambda^E$, $\lambda^D$, $\lambda^Q$
and $\lambda^H$, for consideration while assuming the rest of them,
$\lambda$, $\lambda^{'}$ and $\lambda^{QD}$, are negligible.
The realization through different LPV trilinear coupling
combinations can be derived straightforwardly. The method to
calculate the neutrino mass matrix we use is given in
Appendix C.

Here we list the parameters of neutrino oscillation given by
experiments, $\Delta m^2_{21}=(7.59 \pm 0.21)\times 10^{-5}~
\mathrm{eV}^2,~ \Delta m^2_{32}=(2.43 \pm 0.13)\times 10^{-3}~
\mathrm{eV}^2$ and $\sin^2 2\mathrm{\theta}_{12} =
0.861^{+0.026}_{-0.022}$,~ $\sin^2 2\mathrm{\theta}_{23} > 0.92$,
$\sin^2 2\mathrm{\theta}_{13} = 0.088 \pm 0.008$. Scanning the
parameter space with proper EWSB, we find by adjusting the ratios and
values of the LPV trilinear couplings we choosing, the correct neutrino
spectrum can be generated through
the $\lambda^E\lambda^E$, $\lambda^D \lambda^Q$ and
$\lambda^H \lambda^H$ type one-loop diagrams.
Numerical illustration is shown in Table
I, in Set I we take the mass mixings assumptions mentioned before,
in Set II we take different mass mixings and bigger vector-like
masses for comparison.
The specific parameters settings see Appendix C.
\begin{table}
\begin{tabular}[c]{|c|c|c|c|c|c|}
\hline
\phantom{XXXX} &
\phantom{i}$M^{\nu}_{ij} |_{\lambda^E_{i3} \lambda^E_{j3}} $  \phantom{i}&
\phantom{i}$M^{\nu}_{ij} |_{\lambda^D_{i3} \lambda^Q_{j3}}$ \phantom{i} &
\phantom{i}$M^{\nu}_{ij} |_{\lambda^D_{i3} \lambda^D_{j3}}$ \phantom{i} &
\phantom{i}$M^{\nu}_{ij} |_{\lambda^Q_{i3} \lambda^Q_{j3}}$ \phantom{i} &
\phantom{i}$M^{\nu}_{ij} |_{\lambda^H_i \lambda^H_j}$ \phantom{i}  \\ \hline\hline
$\mathrm{Set}~\mathrm{I}$   &  ~0.0043 &~0.238& 0 &0 & 0.08   \\ \hline
$\mathrm{Set} ~\mathrm{II}$  &~0.0027 &0.168& 0.004  & 0.003 & 0.011 \\ \hline
\end{tabular}
\caption{Numerical illustration for 5 types of one-loop contributions in our model
,the specific parameter settings see Appendix B. $M^{\nu}_{ij}$ ($\mathrm{GeV}$)
stands for the parts in Eq. (4,5) excepting the LPV trilinear coupling constants. }
\end{table}

That is by choosing (for Set I)
\begin{eqnarray}
&&\frac{\lambda^{Q}_{13}}{\lambda^{Q}_{23}}\sim 0.25~,~
\frac{\lambda^{Q}_{33}}{\lambda^{Q}_{23}}\sim 1.4~,~
\lambda^{Q}_{23}\sim 2.1\times 10^{-6}~,~
\lambda^{Q}_{13}\sim\lambda^{D}_{13}~,~
\lambda^{Q}_{23}\sim\lambda^{D}_{23}~,~
\lambda^{Q}_{33}\sim\lambda^{D}_{33}~,~
\nonumber\\
&&~~~~~~~~~~~~~
\frac{\lambda^H_1}{\lambda^H_2}\sim 1.4~,~
\frac{\lambda^H_3}{\lambda^H_2}\sim 1~,~
\lambda^H_2\sim\lambda^{Q}_{23}~,~
\lambda^{E}_{13}\sim\lambda^{E}_{23}\sim\lambda^{E}_{33}\sim\lambda^{Q}_{23}~,~
\end{eqnarray}
we have
\begin{eqnarray}
&&\frac{m_{\nu 2}}{m_{\nu 3} }\sim 0.17~,~
m_{\nu 2}\sim  5.9\times10^{-4}~ \mathrm{eV}~, ~
m_{\nu 1}\sim 5.1\times10^{-6}~ \mathrm{eV}~,~
\nonumber\\
&&~~~~~~
\sin\theta_{13}\sim0.143~, ~\sin\theta_{23}\sim0.581~,~ \sin\theta_{12}\sim0.559~.
\end{eqnarray}

Unlike in the 3G LPV case, where $\lambda'_{i33} \lambda'_{j33}$,
$\lambda'_{i23} \lambda'_{j32}+\lambda'_{i32} \lambda'_{j23}$ and
$\lambda_{i33} \lambda_{j33}$ type one-loop contributions are dominant,
subdominant and next-to-subdominant, here in our model, under the assumptions
mentioned before, $\lambda^Q_{i3} \lambda^D_{j3}$, $\lambda^H_{i} \lambda^H_{j}$ and
$\lambda^E_{i3} \lambda^E_{j3}$ type one-loop contributions are dominant,
subdominant and next-to-subdominant, respectively. This is because the new
fermions $\tau_1,~t_{1,2},~b_{1,2}$ in the internal lines, see Fig. 1,
are much heavier than the third generation fermions $\tau,~t,~b$.
\vspace{2pt}
\begin{figure}[htbp]
\centering
\includegraphics[width=2in]{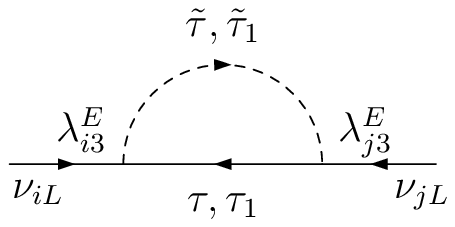}
\includegraphics[width=2in]{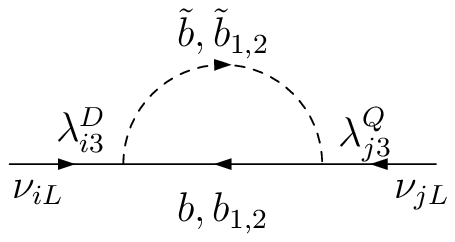}
\includegraphics[width=2in]{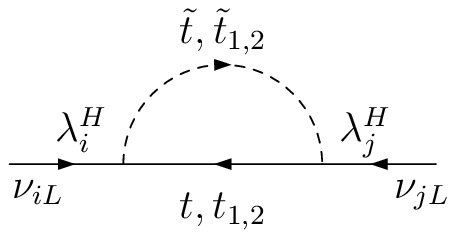}

\caption{New one-loop contributions to the the neutrino masses and mixings
from $\lambda^E \lambda^E$, $\lambda^Q \lambda^D$ and $\lambda^H \lambda^H$
type couplings. All particles stay in mass eigenstates. }
\end{figure}

For the same reason, our requirements of the new LPV couplings we choose
are of order $10^{-6}$ and small enough to avoid measurable FCNC decays
such as $\mu\rightarrow e \gamma$ \cite{FCNC}.
It worth to note that by decoupling the vector-like generation,
correct neutrino masses and mixings cannot be obtained via
$\lambda \lambda$, $\lambda{'} \lambda{'}$ type one-loop contributions.

In addition, $\lambda^H \lambda^H$ type
contribution containing up-type (s)quarks in the internal lines is
absent in 3G LPV models because the vector-like down-type doublet quark
$Q^b_H$ mixe with the right hand singlet top quark.

From Table I, we can also see that by choosing $\lambda^Q_{i3} \lambda^D_{j3}$,
$\lambda^Q_{i3} \lambda^Q_{j3}$ and $\lambda^D_{i3} \lambda^D_{j3}$ type
one-loop contributions, the correct neutrino spectrum can also been
generated in parameters Set II, we don't list the detailed results here.

\section{HIGGS MASS}

There are four new up-type Higgs Yukawa couplings in our model,
$y^U,~y^Q,~y^{QU},~y^{H'}$, corresponding to the Yukawa mass,
$m^t_{34},~m^t_{43},~m^t_{44},~m^b_{H}$, separately, and also five
new down-type Higgs Yukawa couplings, $y^E,~y^D,~y^{Q'},~y^{QD},~y^{H}$,
corresponding to the Yukawa mass,
$m^\tau_{34},~m^b_{34},~m^b_{43},~m^b_{44},~m^t_{H}$, separately. The related
superpotential contributing to the lightest scalar Higgs mass is
shown in ${\mathcal W}_0$.
According to the assumptions mentioned in the last section, we
neglect the down-type Higgs Yukawa contributions and the small
up-type contributions between the SM third generations and the extra
vector-like generations. The relevant superpotential can be
simplified as
\begin{eqnarray}
{\mathcal W}&=& \mu^Q Q_4 Q_H+\mu^D D_4^c D_H^c
          +y^{H'} Q_H H_u D_H^c+ y^{QU} Q_4 H_u U_4^c~,
\end{eqnarray}

So when neglecting the small D-term and the two-loop contribution,
the new one-loop contribution to the lightest scalar Higgs
square-mass is \cite{Higgs1,Higgs2}
\begin{eqnarray}
\bigtriangleup m^2_h=\frac{3\times 2}{4\pi^2}~(y^{t}_V)^4 ~v^2\sin^4\beta~
[t_V-\frac{1}{6}(5-\frac{1}{x})(1-\frac{1}{x})
+2~\frac{X^2_{V}}{M^2_S}(1-\frac{1}{3x})],
\end{eqnarray}
where $v=174\mathrm{Gev}$ indicates the Higgs VEV and
\begin{eqnarray}
&&~~~~~~~~~~~
y^{H'}=y^{QU}\equiv y^{t}_V~,~x=M^2_S/M^2_V~,~
t_V=\log\frac{M^2_S}{M^2_V}~,\\
&&(A_{H'}-\mu_{H'} \cot \beta)^2=(A_{QU}-\mu_{QU} \cot \beta)^2
=(A_V-\mu_V \cot \beta)^2\equiv X^2_{V},\nonumber
\end{eqnarray}
in which, for simplicity, $\mu^Q=\mu^D\equiv M_V$ stands for the vector-like mass
of the new up-type quarks, $M^2_Q=M^2_D\equiv m^2$ (see Eq.(4)) and $M_S=\sqrt{M^2_V+m^2}$
stands for average mass of the new up-type squarks.

In MSSM, the Higgs mass from the $t,~\tilde{t}$ one-loop contributions is
about 110 GeV, for $A^t=\mu=400$ GeV, $m_{\tilde{t}}=400$ GeV and
$\tan \beta=10$. Direct search bounds from CMS for exotic heavy top-like quark set
limits of $M_{t'}>557$ GeV if $B(t'\rightarrow Wb)=1$ \cite{CMS1}
and $M_{t'}>475$ GeV if $B(t'\rightarrow Zt)=1$ \cite{CMS2}. When considering
the mass mixing between the vector-like quarks and the SM third
generation quarks, in other words, considering the realistic branch ratios,
the mass limit is adjust to be $M_{t'}>415$ GeV \cite{bound,martin}.
So if we set the vector-like fermion masses in our model to be
$M_V\sim$ 500 GeV, the soft supersymmetry-breaking parameters to be
$m\sim $ 700 GeV, $A_V =\mu_V\sim500$GeV and $B_V \mu_V\sim
500^2$ GeV$^2$, then from Eq. (10), in order to get approximately
125 GeV Higgs mass, for about $M_V=500$ GeV and $M_S=850$ GeV,
we just need to set $y^{t}_V\sim 1$,  or say, need to set
$m^t_{44}=m^b_{H}\equiv m^{t}_V\sim174$ GeV. These values are just near
their infrared quasi-fix point, as mentioned in last section.

Evoked by the ATLAS and CMS discovery of the enhancement in $\gamma\gamma$
channel and little deviation in ZZ channel \cite{ATLAS, CMS3},
the effects of the exotic vector-like quarks to the Higgs production and decay
have been extensively studied recently \cite{VL}. In general, in a theory with
$N$ vector-like generations extension, the new fermion contributions are suppressed
by $N^2 m^{t2}_V/M^2_V$ \cite{Wise,VL}. So only the very large couplings to the Higgs
can obviously enhance the Higgs production and decay in the $\gamma\gamma$ channel \cite{VL},
but as we have mentioned, these couplings have quasi-fix point which limits their
TeV values to be about 1 \cite{Higgs2}. This value is large enough to accommodate
$m_h\sim 125$ GeV, but too small to influence the Higgs decay, one can't
depend on vector-like fermions by themselves to modify the Higgs decay branching ratios.
As far as the Higgs problem to be concerned, extra vector-like fermions
are mainly introduced to adjust the Higgs mass. However, the $\gamma\gamma$
and ZZ channel anomaly, if they persist, can be realized through the
light stop scenario \cite{light}, which beyond our scope in this paper.

\section{THE EXTRA VECTOR-LIKE FERMION DECAYS }

To be clear, we list the new extra vector-like fermions below:
\begin{eqnarray}
\Psi_E=
\begin{pmatrix}
E^c_H \\ \bar{E}^c_4
\end{pmatrix}
,~
\Psi_Q=
\begin{pmatrix}
Q^{t,b}_4 \\ \bar{Q}^{t,b}_H
\end{pmatrix}
,~
\Psi_U=
\begin{pmatrix}
U^c_{4} \\ U_{H}
\end{pmatrix}
,~
\Psi_D=
\begin{pmatrix}
D^c_{4} \\ D_{H}
\end{pmatrix}
,~
\end{eqnarray}
in which $E^c_H$ mixes with $\tau_L$; $E^c_4$ mixes with $\tau_R$; $Q^b_4, D^c_H$
mixes with $b_L$; $D^c_4, Q^t_H$ mixes with $b_R$, $Q^t_4; U^c_H$ mixes with $t_L$, $U^c_4, Q^b_H$
mixes with $t_R$. These exotic heavy fermions can decay into SM bosons, see Fig. 2,
which will analyze bellow. Our analysis agree with the results given in \cite{Higgs2}.
However the slightly difference comes from their neglect of the contributions
proportional to $s^2_W$ in the vertex of Feynman rules.
\vspace{2pt}
\begin{figure}[htbp]
\centering
\includegraphics[width=2in]{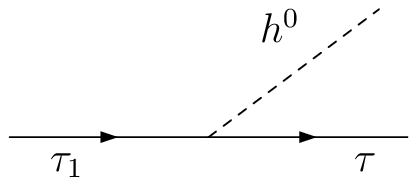}
\includegraphics[width=2in]{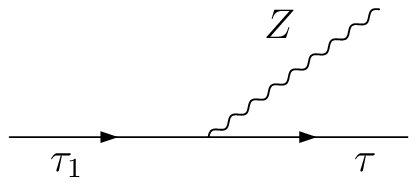}
\includegraphics[width=2in]{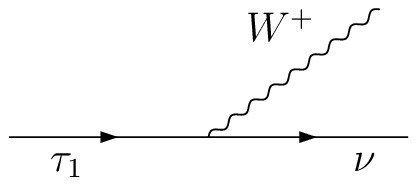}
\includegraphics[width=2in]{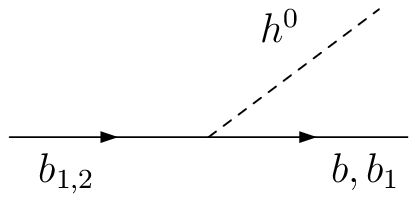}
\includegraphics[width=2in]{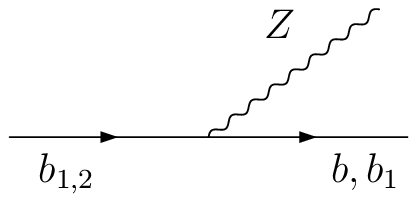}
\includegraphics[width=2in]{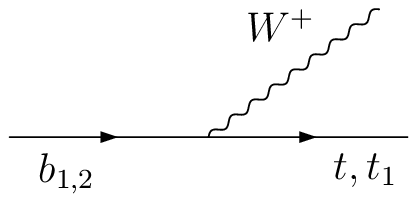}
\includegraphics[width=2in]{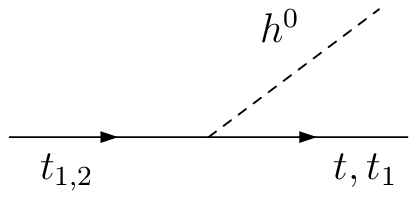}
\includegraphics[width=2in]{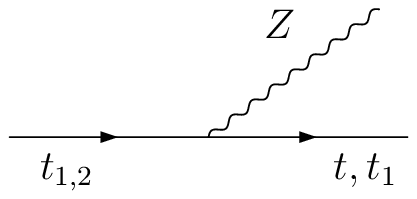}
\includegraphics[width=2in]{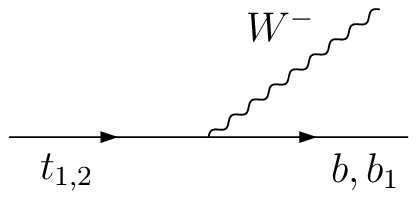}
\caption{Tree-level decay of new exotic fermions in our model, all fermions
stay in mass eigenstates. }
\end{figure}

Note that theoretically speaking, when kinematically allowed,
the exotic fermions predicted in our model have the other two decay modes:
through supersymmetric gauge kinetic interactions or the supersymmetric
Yukawa interactions, decay into chargino/neutralino and sfermions,
such as $\tau_1\rightarrow \tilde{C^+}\tilde{\nu_\tau}$,
$b_1\rightarrow \tilde{N_i}\tilde{b}$, $t_1\rightarrow \tilde{C^-}\tilde{b}$
, where $\tilde{N_i}$, i=1-4, is neutralino, $\tilde{C^{\pm}}$ is chargino;
through LPV interactions, see Eq. (2), decay into fermions and sfermions,
such as $\tau_1\rightarrow e\tilde{\mu}$,
$b_1\rightarrow \tilde{\tau}t$, $t_1\rightarrow \tilde{e}b$.
Although the kinematical conditions for the latter decay mode are easy to be satisfied,
but we have already seen in section III, the LPV couplings in our model, in order to
explain the neutrino spectrum, are of order $10^{-6}$, so we can neglect
this kind of decay channels reasonably. On the other hand, for simplicity
here in our work, we assume the former decay mode is not kinematically allowed.
Therefore, the exotic fermions can only decay into SM bosons.

\subsection{$\tau_{1}$ decays}

The weak bosons interaction Lagrangian to $\tau,~\tau_1$ is
\begin{eqnarray}
{\mathcal L} \supset&& g^W_{\bar{\tau}_{1L}\nu_{\tau}}\bar{\tau}_{1L}\gamma^\mu \nu_{\tau L}W^-_\mu
+g^Z_{\bar{\tau}_{1L}\tau_{L}}\bar{\tau}_{1L}\gamma^\mu \tau_{L}Z_\mu
+g^Z_{\bar{\tau}_{1R}\tau_{R}}\bar{\tau}_{1R}\gamma^\mu \tau_{R}Z_\mu\nonumber\\
&&+g^{h^0}_{\bar{\tau}_{1L}\tau_{R}}\bar{\tau}_{1L} \tau_{R}h^0
+g^{h^0}_{\bar{\tau}_{L}\tau_{1R}}\bar{\tau}_{L} \tau_{1R}h^0
+\mathrm{h.c.},
\end{eqnarray}
the couplings and the decay widths of $\mathrm{\tau_{1}}$ are given in Appendix D.

\vspace{2pt}
\begin{figure}[htbp]
\centering
\includegraphics[width=3in]{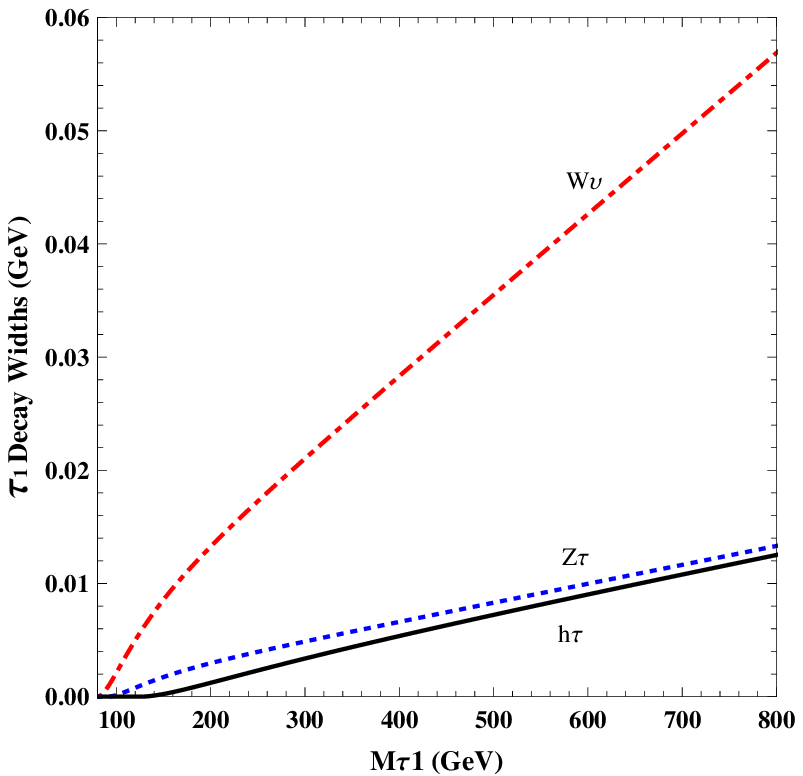}
\includegraphics[width=3in]{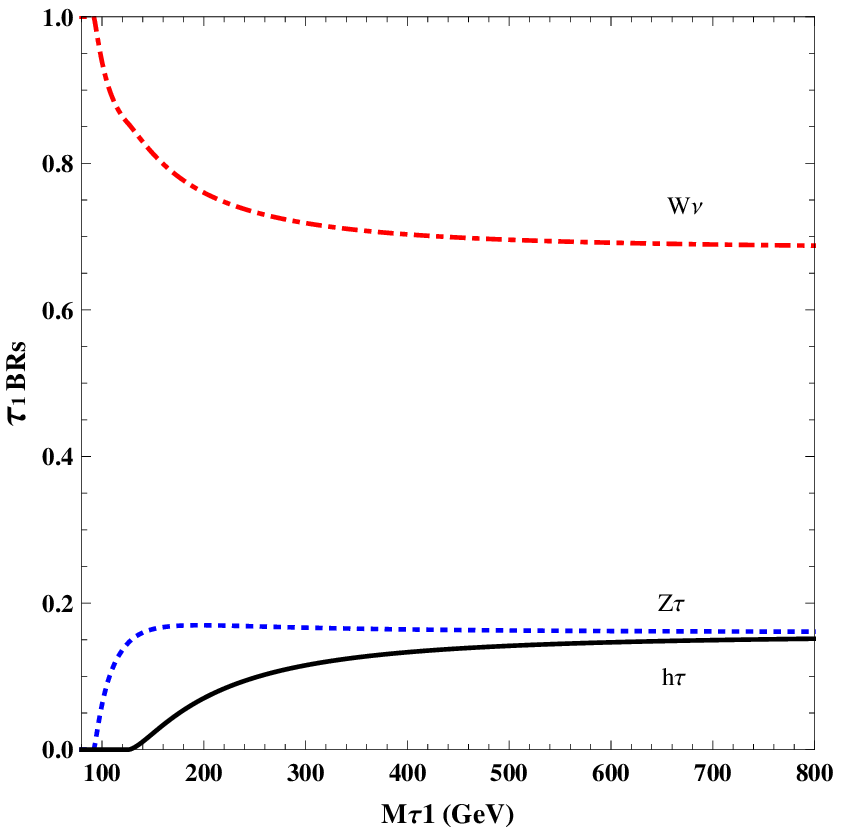}
\caption{The decay widths of the new lepton $\mathrm{\tau_{1}}$ (left panel) and
its branching ratios (right  panel) with $y^E=0.04$. }
\end{figure}

The main characteristic of the lepton sector is that there must be
mass mixing between the third and the vector-like lepton,
otherwise the new heavy charged leptons $\mathrm{\tau_{1}}$ will
be stable and give unacceptable cosmological heavy
relic \cite{stable1}. For Specific, when $y^E=0$, the off-diagonal
elements of $L^\tau$, $R^\tau$ are equal to zero. That's why we
set $y^E\neq0$ in section II while discussing neutrino spectrum,
more specifically, we set $y^E\leq0.04$. Under these parameters
settings, numerical results of $\tau_{1}$ decay into $W, Z, h^0$
are shown in Fig. 3,
we can see in the limit of large $m_{\tau_1}$, the branching rations
are $\mathrm{BR}(\mathrm{\tau_{1}\rightarrow W \nu_{\tau}})\sim0.7$ and
$\mathrm{BR}(\mathrm{\tau_{1}\rightarrow Z \tau})
=\mathrm{BR}(\mathrm{\tau_{1}\rightarrow h \tau})\sim0.15$ .

\subsection{$t_{1,2}$~ decays}

The weak bosons interaction Lagrangian to $t,~t_1,~t_2$ is
\begin{eqnarray}
{\mathcal L} \supset&& g^W_{\bar{t}_{1L}b_{L}}\bar{t}_{1L}\gamma^\mu b_{L}W^-_\mu
+g^W_{\bar{t}_{2L}b_{L}}\bar{t}_{2L}\gamma^\mu b_{L}W^-_\mu
+g^W_{\bar{t}_{1L}b_{R}}\bar{t}_{1R}\gamma^\mu b_{R}W^-_\mu
+g^W_{\bar{t}_{2R}b_{R}}\bar{t}_{2R}\gamma^\mu b_{R}W^-_\mu\nonumber\\
&& g^W_{\bar{t}_{2L}b_{1L}}\bar{t}_{2L}\gamma^\mu b_{1L}W^-_\mu
+g^W_{\bar{t}_{2L}b_{R1}}\bar{t}_{2R}\gamma^\mu b_{1R}W^-_\mu
+g^Z_{\bar{t}_{1L}t_{L}}\bar{t}_{1L}\gamma^\mu t_{L}Z_\mu
+g^Z_{\bar{t}_{2L}t_{L}}\bar{t}_{2L}\gamma^\mu t_{L}Z_\mu\nonumber\\
&&+g^Z_{\bar{t}_{2L}t_{1L}}\bar{t}_{2L}\gamma^\mu t_{1L}Z_\mu
+g^Z_{\bar{t}_{1R}t_{R}}\bar{t}_{1R}\gamma^\mu t_{R}Z_\mu
+g^Z_{\bar{t}_{2R}t_{R}}\bar{t}_{2R}\gamma^\mu t_{R}Z_\mu
+g^Z_{\bar{t}_{2R}t_{1R}}\bar{t}_{2R}\gamma^\mu t_{1R}Z_\mu\nonumber\\
&&+g^{h^0}_{\bar{t}_{1L}t_{R}}\bar{t}_{1L}t_{R}h^0
+g^{h^0}_{\bar{t}_{L}t_{1R}}\bar{t}_{L} t_{1R}h^0
+g^{h^0}_{\bar{t}_{2L}t_{R}}\bar{t}_{2L} t_{R}h^0
+g^{h^0}_{\bar{t}_{L}t_{2R}}\bar{t}_{L} t_{2R}h^0\nonumber\\
&&+g^{h^0}_{\bar{t}_{2L}t_{1R}}\bar{t}_{2L} t_{1R}h^0
+g^{h^0}_{\bar{t}_{1L}t_{2R}}\bar{t}_{1L} t_{2R}h^0
+\mathrm{h.c.},
\end{eqnarray}
the couplings and the decay widths of $\mathrm{t_{1,2}}$ are given in Appendix D.

\vspace{2pt}
\begin{figure}[htbp]
\centering
\includegraphics[width=3in]{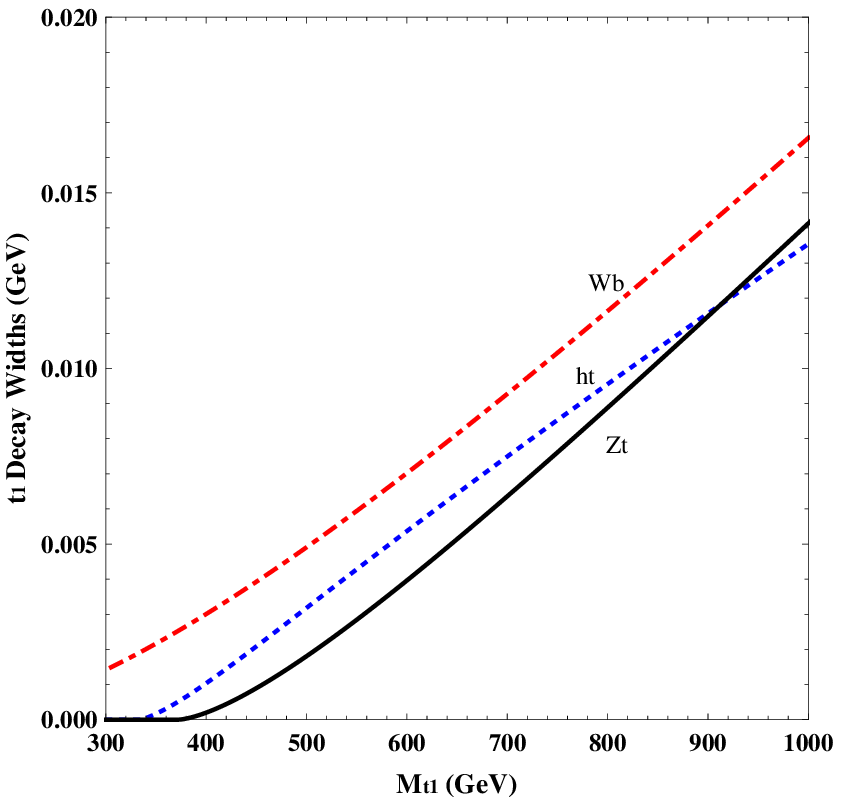}
\includegraphics[width=3in]{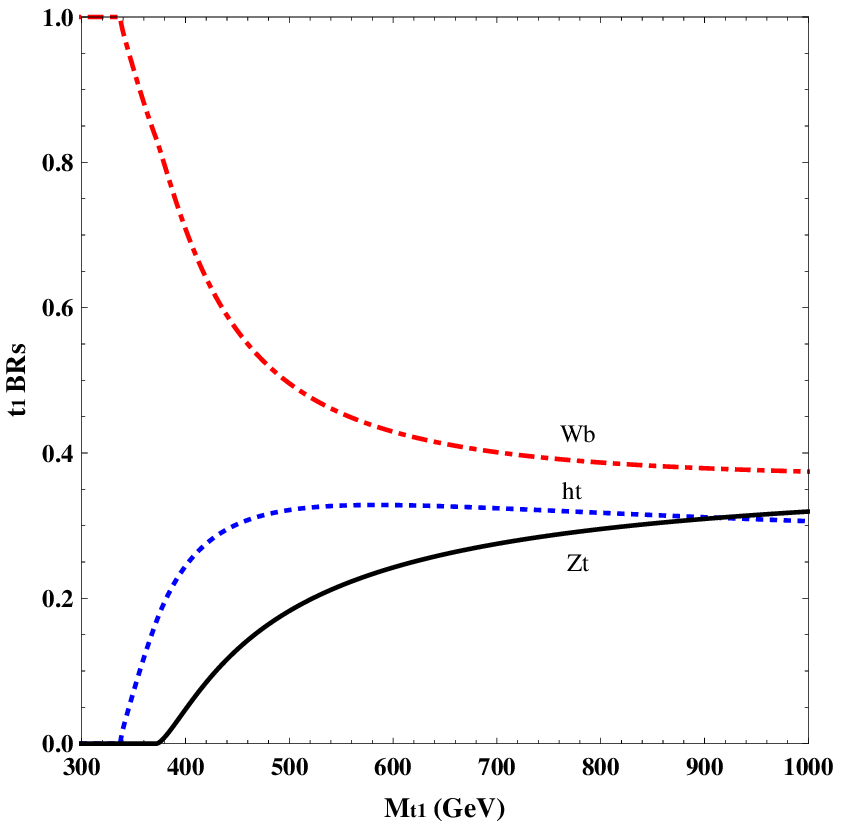}
\caption{The decay widths of the lightest new up-type quark $\mathrm{t_{1}}$ (left panel) and
its branching ratios (right  panel) with $y^{QD}=y^{H}=y^{D}=0$, $y^U\sim y^Q=0.08$
and $y^{QU}\sim y^{H'}=1$. }
\end{figure}
\vspace{2pt}
\begin{figure}[htbp]
\centering
\includegraphics[width=3in]{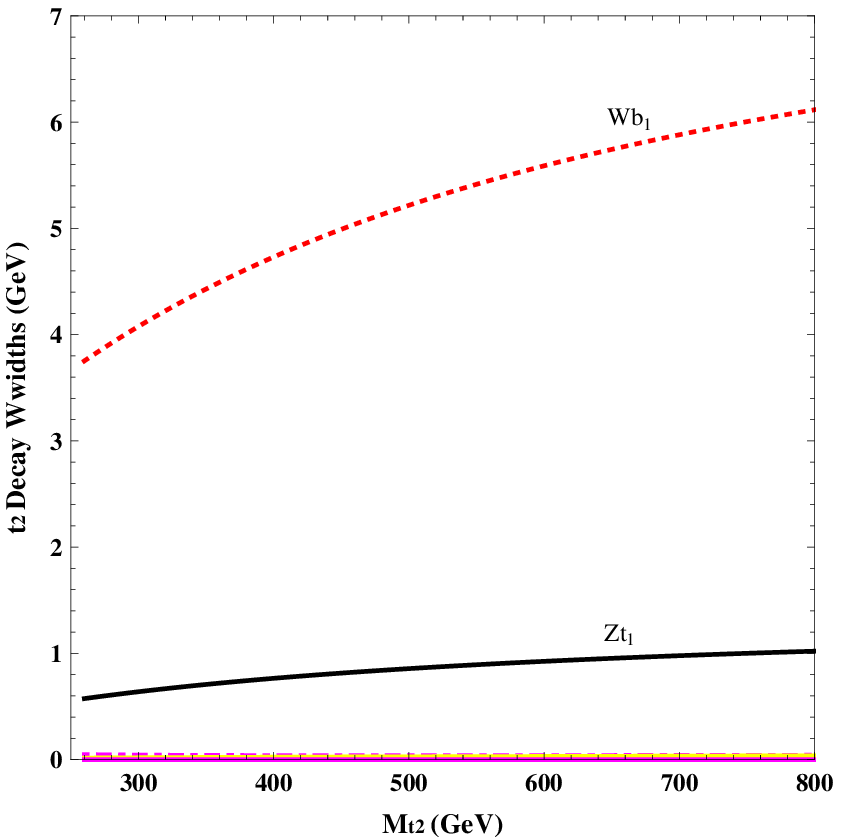}
\includegraphics[width=3in]{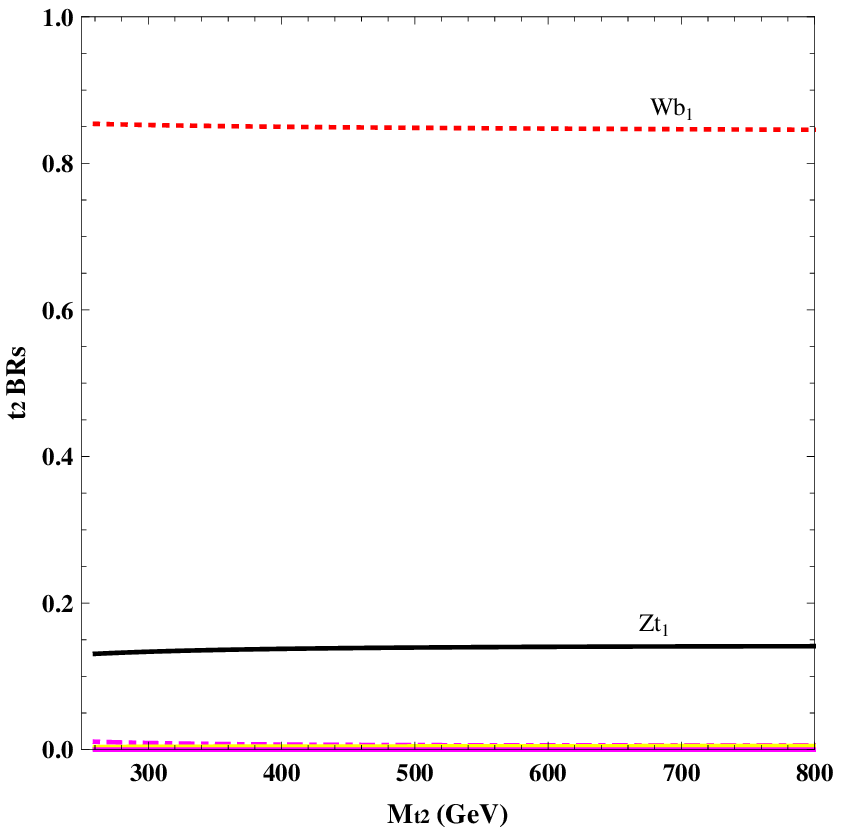}
\caption{The decay widths of the heaviest new up-type quark $\mathrm{t_{2}}$ (left panel) and
its branching ratios (right  panel) with $y^{QD}=y^{H}=y^{D}=0$, $y^U\sim y^Q=0.08$
and $y^{QU}\sim y^{H'}=0.98$. }
\end{figure}

As mentioned in section II, we take $y^U\sim y^Q\leq0.08$,
$y^{QU}\sim y^{H'}\leq1$, the numerical reasults are shown in
Fig 4, 5. We can see in the limit of large $M_{t_{1,2}}$, the branching rations
of $t_1$ are
$\mathrm{BR}(\mathrm{t_{1}\rightarrow W b_{}})\sim0.4$ and
$\mathrm{BR}(\mathrm{t_{1}\rightarrow Z t_{}})
=\mathrm{BR}(\mathrm{t_{1}\rightarrow h^0 t_{}})\sim0.3$,
the branching rations of $t_2$ are
$\mathrm{BR}(\mathrm{t_{2}\rightarrow W b_{1}})\sim0.85$ and
$\mathrm{BR}(\mathrm{t_{2}\rightarrow Z t_{1}})\sim0.15$ .

\subsection{$b_{1,2}$~ decays}

The weak bosons interaction Lagrangian to $b,~b_1,~b_2$ is
\begin{eqnarray}
{\mathcal L} \supset&& g^W_{\bar{b}_{1L}t_{L}}\bar{b}_{1L}\gamma^\mu t_{L}W^-_\mu
+g^W_{\bar{b}_{2L}t_{L}}\bar{b}_{2L}\gamma^\mu t_{L}W^-_\mu
+g^W_{\bar{b}_{1L}t_{R}}\bar{b}_{1R}\gamma^\mu t_{R}W^-_\mu
+g^W_{\bar{b}_{2R}t_{R}}\bar{b}_{2R}\gamma^\mu t_{R}W^-_\mu\nonumber\\
&&+g^W_{\bar{b}_{2L}t_{1L}}\bar{b}_{2L}\gamma^\mu t_{1L}W^-_\mu
+g^W_{\bar{b}_{2R}t_{R1}}\bar{b}_{2R}\gamma^\mu t_{1R}W^-_\mu
+g^Z_{\bar{b}_{1L}b_{L}}\bar{b}_{1L}\gamma^\mu b_{L}Z_\mu
+g^Z_{\bar{b}_{2L}b_{L}}\bar{b}_{2L}\gamma^\mu b_{L}Z_\mu\nonumber\\
&&+g^Z_{\bar{b}_{2L}b_{1L}}\bar{b}_{2L}\gamma^\mu b_{1L}Z_\mu
+g^Z_{\bar{b}_{1R}b_{R}}\bar{b}_{1R}\gamma^\mu b_{R}Z_\mu
+g^Z_{\bar{b}_{2R}b_{R}}\bar{b}_{2R}\gamma^\mu b_{R}Z_\mu
+g^Z_{\bar{b}_{2R}b_{1R}}\bar{b}_{2R}\gamma^\mu b_{1R}Z_\mu\nonumber\\
&&+g^{h^0}_{\bar{b}_{1L}b_{R}}\bar{b}_{1L} b_{R}h^0
+g^{h^0}_{\bar{b}_{L}b_{1R}}\bar{b}_{L} b_{1R}h^0
+g^{h^0}_{\bar{b}_{2L}b_{R}}\bar{b}_{2L} b_{R}h^0
+g^{h^0}_{\bar{b}_{L}b_{2R}}\bar{b}_{L} b_{2R}h^0\nonumber\\
&&+g^{h^0}_{\bar{b}_{2L}b_{1R}}\bar{b}_{2L} b_{1R}h^0
+g^{h^0}_{\bar{b}_{1L}b_{2R}}\bar{b}_{1L} b_{2R}h^0
+\mathrm{h.c.}
\end{eqnarray}
the couplings and the decay widths of $\mathrm{b_{1,2}}$ are given in Appendix D.

\vspace{2pt}
\begin{figure}[htbp]
\centering
\includegraphics[width=3in]{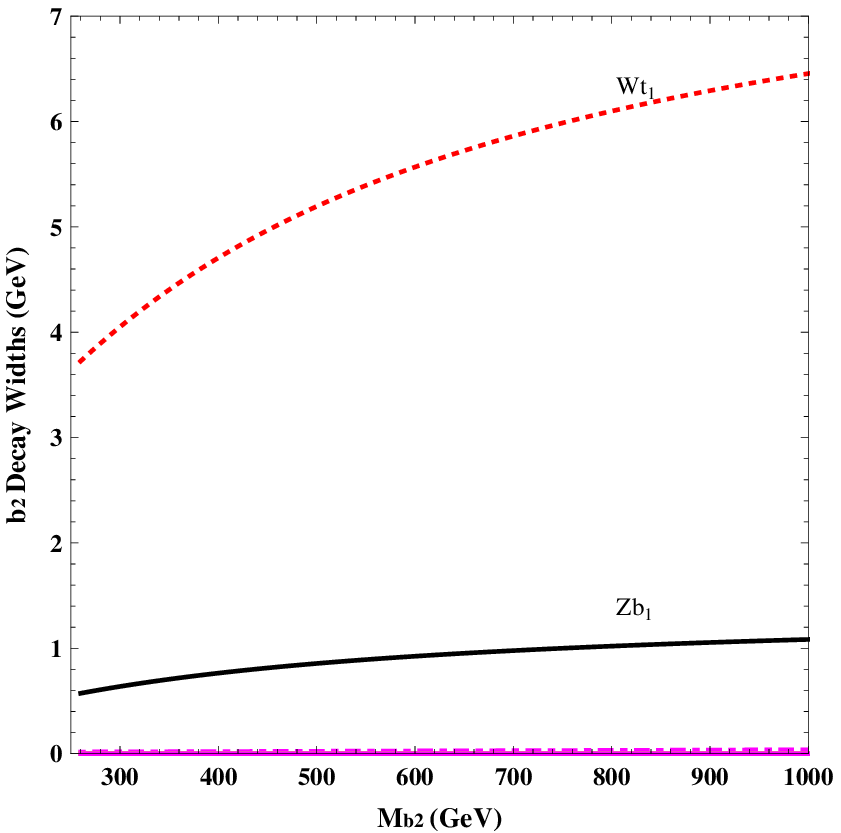}
\includegraphics[width=3in]{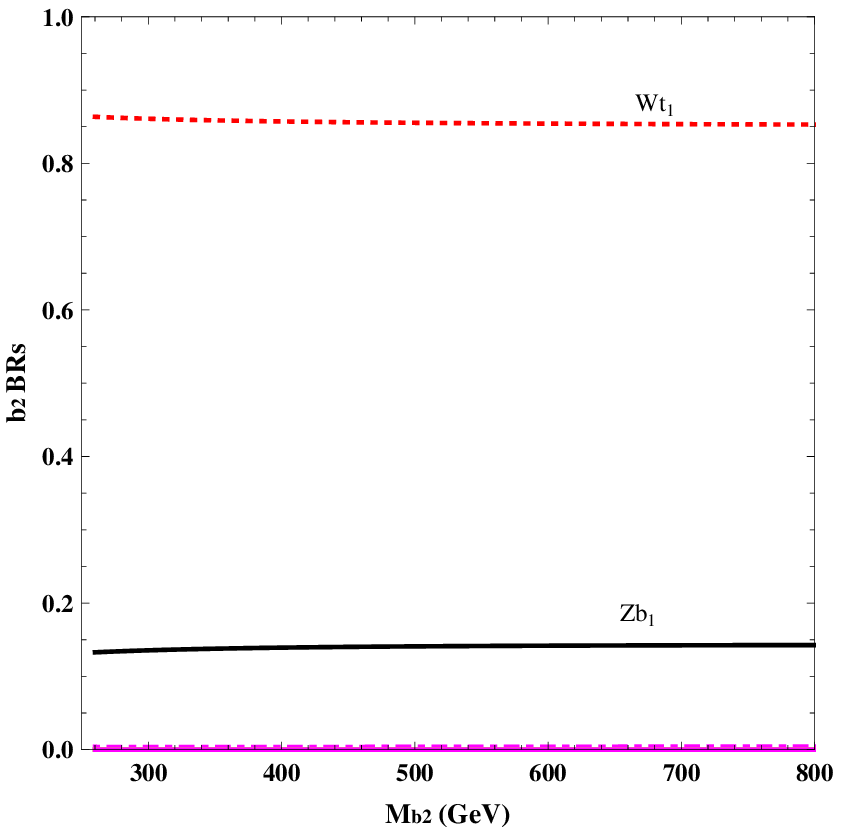}
\caption{The decay widths of the heaviest new down-type quark $\mathrm{b_{2}}$ (left panel) and
its branching ratios (right  panel) with $y^{QD}=y^{H}=y^{D}=0$, $y^U\sim y^Q=0.08$
and $y^{QU}\sim y^{H'}=0.98$.  }
\end{figure}

The numerical results of $\mathrm{b_{2}}$ decay widths and branching rations
are shown in Fig. 6 under the parameter settings mentioned before, we can see
$\mathrm{BR}(\mathrm{b_{2}\rightarrow W b_{1}})\sim0.85$ and
$\mathrm{BR}(\mathrm{b_{2}\rightarrow Z b_{1}})\sim0.15$.
The branching rations of $\mathrm{b_{1}}$ are
$\mathrm{BR}(\mathrm{b_{1}\rightarrow
W t})=1$ which are not shown here.

\section{summary and discussion}

We have studied several phenomenological aspects of the LPV MSSM model with
a vector-like extra generation: the neutrino spectrum, the Higgs mass and the
LHC phenomenology of the new predicted fermions. The results are:
\begin{itemize}
    \item
        The correct neutrino masses and mixings, especially the
        relatively large $\theta_{13}$ can be generated from
        trilinear LPV couplings.  The new trilinear R-parity violating
        couplings make it easy to generate the proper value of
        $\theta_{13}$.  These coupling constants need to be about
        $10^{-6}$.
    \item
        The two new up-type Higgs Yukawa couplings, $y^{H'}$ and
        $y^{QU}$, between the vector-like quarks and the SM third generation
        quarks, with values about 1 near to their infrared quasi-fixed point
        in TeV scale, can give rise to 125 GeV Higgs mass with no need
        of very heavy new superpartner.
    \item
        There are five new heavy fermions, $\mathrm{\tau}_1$,
        $\mathrm{t}_{1,2}$, $\mathrm{b}_{1,2}$ , predicted in this model.
        They can only decay into SM bosons by some kinematic assumptions.
        The branching radio depend on the mass mixing
        between the vector-like fermions and the SM third generation fermions.
        These charged exotic fermions would be quasi-stable if such mass mixings are
        very small.
\end{itemize}

Based on our previous work about bilinear LPV couplings, further
research on the renormalization group (RG) of them is worthy to be
studied in the future. There are also plenty of aspects to be further
analyzed in the area of new fermion LHC phenomenology based on
this model.

\begin{acknowledgments}
We would like to thank Dr. Guang-Zhi Xu for a very helpful discussion.
This work was supported in part by the National Natural Science
Foundation of China under nos.11075193 and 10821504, and by the National
Basic Research Program of China under Grant No. 2010CB833000.
\end{acknowledgments}

\appendix{}

\section{THE (S)FERMION MASS MIXING MATRIXES}

Because the CP violation is not considered in this paper, we have taken all the masses
real. In this model, the mass matrix M of the third generation lepton and the vector-like
lepton is given as following
\begin{equation}
{\mathcal L} \supset -\left(\tau, E_H^c\right) {\mathcal M}^\tau
\begin{pmatrix}
\tau^c \\ E_4^c
\end{pmatrix}
,
\end{equation}
and
\begin{equation}
{\mathcal M}^l =
\begin{pmatrix}
m_{33}^\tau & m_{34}^\tau \\
0        & \mu^E
\end{pmatrix}
,
\end{equation}
where $m_{33}^\tau\equiv y_{33}^l\displaystyle\frac{v}{\sqrt{2}}\cos\beta$
and $m_{34}^\tau\equiv y_3^E\displaystyle\frac{v}{\sqrt{2}}\cos\beta$. Taking
$|\mu^E|\gg|m_{34}^\tau|$, then the biunitary matrix to
diagonalize ${\mathcal M}^\tau$ are
\begin{equation}
L^{\tau*}{\mathcal M}^\tau R^{\tau\dagger}=(m_\tau,~\mu^E)\equiv \mathrm{diag}(m_{\tau},~m_{\tau1})~,
\end{equation}
where
\begin{equation}
L^{\tau} =
\begin{pmatrix}
1 & \frac{m_{34}^\tau}{\mu^E} \\
-\frac{m_{34}^\tau}{\mu^E} & 1  \\
\end{pmatrix}
,~
R^{\tau} =
\begin{pmatrix}
1 & \frac{m_\tau m_{34}^\tau}{(\mu^E)^2} \\
-\frac{m_\tau m_{34}^\tau}{(\mu^E)^2} & 1 \\
\end{pmatrix}
.
\end{equation}

The mass matrix ${\mathcal M}^b$ of the third generation down-quark and
vector-like down-type quarks is given as following
\begin{equation}
{\mathcal L} \supset -\left(b, Q_4^b, D_H^c\right) {\mathcal M}^b
\begin{pmatrix}
b^c \\ D_4^c \\ Q_H^t
\end{pmatrix}
,
\end{equation}
where
\begin{equation}
{\mathcal M}^b =
\begin{pmatrix}
m_{33}^b & m_{34}^b & 0      \\
m_{43}^b & m_{44}^b & \mu^Q \\
0        & \mu^D   & m_H^b
\end{pmatrix}
,
\end{equation}
where
$m_{33}^b\equiv y_{33}^d\displaystyle\frac{v}{\sqrt{2}}\cos\beta$,
$m_{34}^b\equiv y_3^D\displaystyle\frac{v}{\sqrt{2}}\cos\beta$,
$m_{43}^b\equiv y_3^{Q\prime}\displaystyle\frac{v}{\sqrt{2}}\cos\beta$ and
$m_{44}^b\equiv y^{QD}\displaystyle\frac{v}{\sqrt{2}}\cos\beta$.
Taking that
$|\mu^Q|\sim|\mu^D|\gg|m_{4b}^b|,~|m_{44}^b|,~|m_{33}^b|,~|m_{34}^b|$,
then the biunitary matrix diagonalize ${\mathcal M}^t$ are
\begin{equation}
L^{b*}{\mathcal M}^b R^{b\dagger}=(m_b,~\mu^Q,~\mu^D)\equiv \mathrm{diag}(m_{b},~m_{b1},~m_{b2})~,
\end{equation}
where
\begin{equation}
L^{b} =
\begin{pmatrix}
1 & 0 & -\frac{m^b_{34}}{\mu^D}  \\
0 & 1 & \frac{\mu^D m_{44}^t+\mu^Q m_H^b}{(\mu^{D})^2+(\mu^{Q})^2-(m_H^{b})^2} \\
\frac{m^b_{34}}{\mu^D} & \frac{(m_H^{b})^2+(m^b_{34})^2}{\mu^{Q} m_H^{b}+\mu^D m_{44}^b }  & 1
\end{pmatrix}
,
\end{equation}
and
\begin{equation}
R^{b} =
\begin{pmatrix}
1 & \frac{m^b_{34}}{\mu^D} & 0  \\
0 & \frac{\mu^Q m_{44}^b+\mu^D m_H^b}{(\mu^{D})^2+(\mu^{Q})^2-(m_H^{b})^2} & 1 \\
-\frac{m^b_{34}}{\mu^D} & 1  & \frac{(m_H^{b})^2+(m^b_{34})^2}{\mu^{D} m_H^{b}+\mu^Q m_{44}^b }
\end{pmatrix}
.
\end{equation}

The mass matrix ${\mathcal M}^t$ of the top quark and vector-like
up-type generations is given as following
\begin{equation}
{\mathcal L} \supset -\left(t, ~Q_4^t, ~U_H^c\right) {\mathcal M}^t
\begin{pmatrix}
t^c \\ U_4^c \\ Q_H^b
\end{pmatrix}
,
\end{equation}
where
\begin{equation}
{\mathcal M}^t =
\begin{pmatrix}
m_{33}^t & m_{34}^t & 0      \\
m_{43}^t & m_{44}^t & \mu^Q \\
0        & \mu^U    & m_H^t
\end{pmatrix}
,
\end{equation}
where $m_{33}^t\equiv y_{33}^u\displaystyle\frac{v}{\sqrt{2}}\sin\beta$,
$m_{34}^t\equiv y_3^U\displaystyle\frac{v}{\sqrt{2}}\sin\beta$,
$m_{43}^t\equiv y_3^Q\displaystyle\frac{v}{\sqrt{2}}\sin\beta$,
$m_{44}^t\equiv y^{QU}\displaystyle\frac{v}{\sqrt{2}}\sin\beta$ and
$m_H^t\equiv y\displaystyle\frac{v}{\sqrt{2}}\cos\beta$.  Taking that
$|\mu^Q|\sim|\mu^U|\gg|m_{43}^t|,~|m_{44}^t|,~|m_{33}^t|,~|m_{34}^t|,~
|m_H^t|$, then the biunitary matrix diagonalize ${\mathcal M}^t$
are
\begin{equation}
L^{t*}{\mathcal M}^t R^{t\dagger}=(m_t,~\mu^Q,~\mu^U)\equiv \mathrm{diag}(m_{t},~m_{t1},~m_{t2})~,
\end{equation}
where
\begin{equation}
L^{t} =
\begin{pmatrix}
1 & 0 & -\frac{m^t_{34}}{\mu^U}  \\
0 & 1 & \frac{\mu^U m_{44}^t+\mu^Q m_H^t}{(\mu^{U})^2+(\mu^{Q})^2-(m_H^{t})^2} \\
\frac{m^t_{34}}{\mu^U} & \frac{(m_H^{t})^2+(m^t_{34})^2}{\mu^{Q} m_H^{t}+\mu^U m_{44}^t }  & 1
\end{pmatrix}
,
\end{equation}
and
\begin{equation}
R^{t} =
\begin{pmatrix}
1 & \frac{m^t_{34}}{\mu^U} & 0  \\
0 & \frac{\mu^Q m_{44}^t+\mu^U m_H^t}{(\mu^{U})^2+(\mu^{Q})^2-(m_H^{t})^2} & 1 \\
-\frac{m^t_{34}}{\mu^U} & 1  & \frac{(m_H^{t})^2+(m^t_{34})^2}{\mu^{U} m_H^{t}+\mu^Q m_{44}^t }
\end{pmatrix}
,
\end{equation}

The charged slepton mass-squared matrix $\tilde{\cal M}_\tau^2$
of $\tilde{\tau}$ and the superpartners of the vector-like leptons is given as
following
\begin{equation}
{\mathcal L} \supset \left(\tilde{L}_3^{-*}, \tilde{E}^c_3,\tilde{E}^c_4,
\tilde{E}_H^{c*}\right) \tilde{\mathcal M}_l^2
\left(\begin{array}{c}
\tilde{L}_3^- \\ \tilde{E}_3^{c*} \\ \tilde{E}_4^{c*} \\ \tilde{E}_H^c
\end{array}\right)\,,
\end{equation}
where
\begin{equation}
\begin{array}{cl}
(\tilde{\mathcal M}_\tau^2)_{11}& =
M^2+\left(\frac{m_Z^2}{2}-m_W^2\right)\cos2\beta+m_\tau^2+|m^\tau_{34}|^2~,~
(\tilde{\mathcal M}_\tau^2)_{12} = (m_0-\mu\tan\beta)m_\tau\,, \\
(\tilde{\mathcal M}_\tau^2)_{13}& = (m_0-\mu\tan\beta)m^{\tau}_{34}~,~
(\tilde{\mathcal M}_\tau^2)_{14} = \mu^e m^{\tau}_{34}~,~
(\tilde{\mathcal M}_\tau^2)_{21} = (m_0-\mu\tan\beta)m_\tau\,, \\
(\tilde{\mathcal M}_\tau^2)_{22}& =
M_E^2-(m_Z^2-m_W^2)\cos 2\beta+m_\tau^2~,~
(\tilde{\mathcal M}_\tau^2)_{23} = m_\tau m^{\tau}_{34}\,,\\
(\tilde{\mathcal M}_\tau^2)_{24}& = 0~,~
(\tilde{\mathcal M}_\tau^2)_{31} = (m_0-\mu\tan\beta)m^\tau_{34}~,~
(\tilde{\mathcal M}_\tau^2)_{32} = m_\tau m^\tau_{34}\,,\\
(\tilde{\mathcal M}_\tau^2)_{33}& =
|\mu^E|^2+M_E^2-(m_Z^2-m_W^2)\cos 2\beta+|m^\tau_{34}|^2~,~
(\tilde{\mathcal M}_\tau^2)_{34} = B^E\mu^E\,, \\
(\tilde{\mathcal M}_\tau^2)_{41}& = \mu^{E} m^\tau_{34}~,~
(\tilde{\mathcal M}_\tau^2)_{42} = 0~,~
(\tilde{\mathcal M}_\tau^2)_{43} = B^{E}\mu^{E},\\
(\tilde{\mathcal M}_\tau^2)_{44} &=
|\mu^E|^2 + M_{EH}^2 + (m_Z^2-m_W^2)\cos 2\beta\,.
\end{array}
\end{equation}
The corresponding unitary scalar matrix is defined as
\begin{equation}
V^{\tau}\tilde{\cal M}^2_\tau V^{\tau\dagger}=
\mathrm{diag}(\tilde{M}^2_{\tau},~\tilde{M}^2_{\tau1},~\tilde{M}^2_{\tau2},~\tilde{M}^2_{\tau3})~,
\end{equation}

The mass-squared matrix $\tilde{\cal M}_b^2$ of $\tilde{b}$
and the superpartners of the down-type vector-like fermions is given
as following
\begin{equation}
{\mathcal L} \supset \left(\tilde{b}^*,\tilde{D}^c_3,\tilde{D}^c_4,
\tilde{D}_H^{c*}, \tilde{Q}_4^{b*}, \tilde{Q}_H^t \right)
\tilde{\mathcal M}_b^2
\left(\begin{array}{c}
\tilde{b} \\ \tilde{D}^{c*}_3\\ \tilde{D}^{c*}_4 \\ \tilde{D}_H^c\\
\tilde{Q}_4^b \\ \tilde{Q}_H^{t*}
\end{array}\right)\,,
\end{equation}
where
\begin{equation}
\begin{array}{cl}
(\tilde{\mathcal M}_b^2)_{11}& =
M_Q^2-\frac{m_Z^2+2m_W^2}{6}\cos 2\beta+m_b^2+|m^b_{34}|^2~,~
(\tilde{\mathcal M}_b^2)_{12} = (m_0-\mu\tan\beta)m_b\,, \\
(\tilde{\mathcal M}_b^2)_{13}& = (m_0-\mu\tan\beta)m^{b}_{34}~,~
(\tilde{\mathcal M}_b^2)_{14} = \mu^D m^{b}_{34}~,~
(\tilde{\mathcal M}_b^2)_{15} = m_bm^b_{43}+m^{b}_{34}m^b_{44}\,, \\
(\tilde{\mathcal M}_b^2)_{16}& =  0~,~
(\tilde{\mathcal M}_b^2)_{21} =(m_0-\mu\tan\beta)m_b~,~
(\tilde{\mathcal M}_b^2)_{22} =
M_D^2-\frac{m_Z^2-m_W^2}{3}\cos 2\beta+m_b^2+|m^d_{43}|^2\,, \\
(\tilde{\mathcal M}_b^2)_{23}& = m_bm^{b}_{34}+m^b_{43}m^{b*}_{44}~,~
(\tilde{\mathcal M}_b^2)_{24} = 0 ~,~
(\tilde{\mathcal M}_b^2)_{25} = (m_0-\mu\tan\beta)m^b_{43}~,~
(\tilde{\mathcal M}_b^2)_{26} = \mu^{Q}m^b_{43}\,, \\
(\tilde{\mathcal M}_b^2)_{31}& =(m_0-\mu\tan\beta)m^b_{34}~,~
(\tilde{\mathcal M}_b^2)_{32} = m_bm^b_{34}+m^{b}_{43}m^b_{44}\,, \\
(\tilde{\mathcal M}_b^2)_{33} &=
|\mu^D|^2+M_D^2-\frac{m_Z^2-m_W^2}{3}\cos2\beta+|m^b_{34}|^2+|m^b_{44}|^2~,~
(\tilde{\mathcal M}_b^2)_{34} = -B^D\mu^D\,, \\
(\tilde{\mathcal M}_b^2)_{35}& = (m_0-\mu\tan\beta)m^b_{44}~,~
(\tilde{\mathcal M}_b^2)_{36} = \mu^{Q}m^b_{44}+\mu^D m_H^{b}~,~
(\tilde{\mathcal M}_b^2)_{41} = \mu^{D}m^b_{34}~,~
(\tilde{\mathcal M}_b^2)_{42} = 0\,, \\
(\tilde{\mathcal M}_b^2)_{43}& = -B^{D}\mu^{D}~,~
(\tilde{\mathcal M}_b^2)_{44} =
|\mu^D|^2+M_{DH}^2+\frac{m_Z^2-m_W^2}{3}\cos 2\beta++|m_H^b|^2\,, \\
(\tilde{\mathcal M}_b^2)_{45}& = \mu^{D}m^b_{44}+\mu^{Q}m_H^b~,~
(\tilde{\mathcal M}_b^2)_{46} = (m_0-\mu\cot\beta)m_H^{b}~,~
(\tilde{\mathcal M}_b^2)_{51} = m_bm^{d}_{43}+m^b_{34}m^{b}_{44}\,, \\
(\tilde{\mathcal M}_b^2)_{52}&= (m_0-\mu\tan\beta)m^{b}_{43}~,~
(\tilde{\mathcal M}_b^2)_{53} = (m_0-\mu\tan\beta)m^{b}_{44}~,~
(\tilde{\mathcal M}_b^2)_{54} = \mu^Dm^{b}_{44}+\mu^{Q}m_H^b\,, \\
(\tilde{\mathcal M}_b^2)_{55}& =
|\mu^Q|^2+M_Q^2-\frac{m_Z^2+2m_W^2}{6}\cos2\beta+|m^b_{44}|^2+|m^b_{43}|^2
~,~
(\tilde{\mathcal M}_b^2)_{56} = B^{Q}\mu^{Q}\,, \\
(\tilde{\mathcal M}_b^2)_{61}& = 0~,~
(\tilde{\mathcal M}_b^2)_{62} = \mu^Qm^{b}_{43}~,~
(\tilde{\mathcal M}_b^2)_{63} = \mu^Qm^{b}_{44}+\mu^D m_H^{b}~,~
(\tilde{\mathcal M}_b^2)_{64} = (m_0-\mu\cot\beta)m_H^{b}\,, \\
(\tilde{\mathcal M}_b^2)_{65}& = B^Q\mu^Q~,~
(\tilde{\mathcal M}_b^2)_{66} =
|\mu^Q|^2+M_{QH}^2++|m_H^b|^2+\frac{m_Z^2+2m_W^2}{6}\cos 2\beta\,.
\end{array}
\end{equation}
The corresponding unitary scalar matrix is defined as
\begin{equation}
V^{b}\tilde{\cal M}^2_b V^{b\dagger}=\mathrm{diag}
(\tilde{M}^2_{b},~\tilde{M}^2_{b1},~\tilde{M}^2_{b2},~\tilde{M}^2_{b3},~\tilde{M}^2_{b4},~\tilde{M}^2_{b5})~,
\end{equation}

The mass-squared matrix $\tilde{\cal M}_t^2$ of $\tilde{t}$
and the superpartners of the up-type vector-like fermions is given as
following
\begin{equation}
{\mathcal L} \supset \left(\tilde{t}^*,\tilde{U}^c_3,\tilde{U}^c_4,
\tilde{U}_H^{c*}, \tilde{Q}_4^{t*}, \tilde{Q}_H^b \right)
\tilde{\mathcal M}_t^2
\left(\begin{array}{c}
\tilde{t} \\ \tilde{U}^{c*}_3 \\ \tilde{U}^{c*}_4\\ \tilde{U}_H^c\\
\tilde{Q}_4^t \\ \tilde{Q}_H^{b*}
\end{array}\right)\,,
\end{equation}
where
\begin{equation}
\begin{array}{cl}
(\tilde{\mathcal M}_t^2)_{11}& =
M_Q^2+\frac{4m_W^2-m_Z^2}{6}\cos 2\beta+m_t^2+|m^t_{34}|^2~,~
(\tilde{\mathcal M}_t^2)_{12} = (m_0-\mu\cot\beta)m_t\,, \\
(\tilde{\mathcal M}_t^2)_{13}& = (m_0-\mu\cot\beta)m^{t}_{34}~,~
(\tilde{\mathcal M}_t^2)_{14} = \mu^U m^{t}_{34} ~,~
(\tilde{\mathcal M}_t^2)_{15} = m_tm^t_{43}+m^{t}_{34}m^t_{44}\,, \\
(\tilde{\mathcal M}_t^2)_{16}& = 0~,~
(\tilde{\mathcal M}_t^2)_{21} = (m_0-\mu\cot\beta)m_t~,~
(\tilde{\mathcal M}_t^2)_{22} =
M_U^2 + \frac{2}{3}(m_Z^2-m_W^2)\cos 2\beta + m_t^2 + |m^t_{43}|^2\,, \\
(\tilde{\mathcal M}_t^2)_{23}& = m_tm^{t}_{34}+m^t_{43}m^{t}_{44}~,~
(\tilde{\mathcal M}_t^2)_{24} = 0~,~
(\tilde{\mathcal M}_t^2)_{25} = (m_0-\mu\cot\beta)m^t_{43}\,, \\
(\tilde{\mathcal M}_t^2)_{26}& = \mu^{Q}m^t_{43}~,~
(\tilde{\mathcal M}_t^2)_{31} = (m_0-\mu\cot\beta)m^t_{34}~,~
(\tilde{\mathcal M}_t^2)_{32}= m_t^*m^u_{34}+m^{t}_{43}m^t_{44}\,, \\
(\tilde{\mathcal M}_t^2)_{33}& =
|\mu^U|^2+M_U^2+\frac{2}{3}(m_Z^2-m_W^2)\cos2\beta+|m^t_{34}|^2+|m^t_{44}|^2
~,~
(\tilde{\mathcal M}_t^2)_{34} = -B^U\mu^U\,, \\
(\tilde{\mathcal M}_t^2)_{35}& = (m_0-\mu\cot\beta)m^t_{44}~,~
(\tilde{\mathcal M}_t^2)_{36} = \mu^{Q}m^t_{44}+\mu^U m_H^{t}~,~
(\tilde{\mathcal M}_t^2)_{41} = \mu^{U}m^t_{34}\,, \\
(\tilde{\mathcal M}_t^2)_{42}& = 0~,~
(\tilde{\mathcal M}_t^2)_{43} = -B^{U}\mu^{U}~,~
(\tilde{\mathcal M}_t^2)_{44} =
|\mu^U|^2 + M_{UH}^2 - \frac{2}{3}(m_Z^2-m_W^2)\cos2\beta+|m_H^t|^2\,,\\
(\tilde{\mathcal M}_t^2)_{45}& = \mu^{U*}m^t_{44} + \mu^Qm_H^{t}~,~
(\tilde{\mathcal M}_t^2)_{46} = (m_0-\mu\tan\beta)m_H^{t}~,~
(\tilde{\mathcal M}_t^2)_{51} = m_tm^{t}_{43}+m^t_{34}m^{t}_{44}\,,\\
(\tilde{\mathcal M}_t^2)_{52}& = (m_0-\mu\cot\beta)m^{t}_{43}~,~
(\tilde{\mathcal M}_t^2)_{53} = (m_0-\mu\cot\beta)m^{t}_{44}~,~
(\tilde{\mathcal M}_t^2)_{54} = \mu^Um^{t}_{44}+\mu^{Q}m_H^t\,, \\
(\tilde{\mathcal M}_t^2)_{55}& =
|\mu^Q|^2+M_Q^2+\frac{4m_W^2-m_Z^2}{6}\cos2\beta+|m^t_{44}|^2+|m^t_{43}|^2
~,~
(\tilde{\mathcal M}_t^2)_{56} = -B^{Q}\mu^{Q}\,, \\
(\tilde{\mathcal M}_t^2)_{61}& = ~,~
(\tilde{\mathcal M}_t^2)_{62} = \mu^Qm^{t}_{43}~,~
(\tilde{\mathcal M}_t^2)_{63} = \mu^Qm^{t}_{44}+\mu^{U}m_H^t~,~
(\tilde{\mathcal M}_t^2)_{64} = (m_0-\mu\tan\beta)m_H^t\,, \\
(\tilde{\mathcal M}_t^2)_{65}& = -B^Q\mu^Q~,~
(\tilde{\mathcal M}_t^2)_{66} =
|\mu^Q|^2+M_{QH}^2+|m_H^t|^2-\frac{4m_W^2-m_Z^2}{6}\cos 2\beta\,.
\end{array}
\end{equation}
The corresponding unitary scalar matrix is defined as
\begin{equation}
V^{t}\tilde{\cal M}^2_t V^{t\dagger}=\mathrm{diag}
(\tilde{M}^2_{t},~\tilde{M}^2_{t1},~\tilde{M}^2_{t2},~\tilde{M}^2_{t3},~\tilde{M}^2_{t4},~\tilde{M}^2_{t5})~,
\end{equation}

\section{NEUTRINO MASSES IN OUR MODEL}

All fourteen types of one loop Fyenman diagrams which can contribute
to the neutrino mass and mixing in our model are shown in Fig. 1
\vspace{2pt}
\begin{figure}[htbp]
\centering
\includegraphics[width=2in]{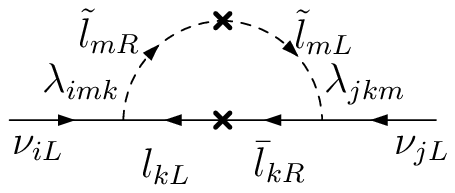}~~~~
\includegraphics[width=2in]{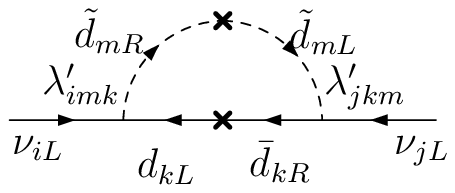}~~~~
\includegraphics[width=2in]{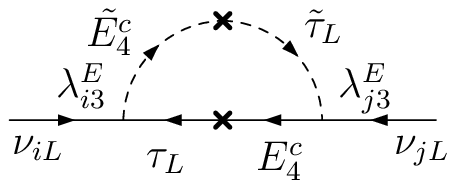}\\
\includegraphics[width=2in]{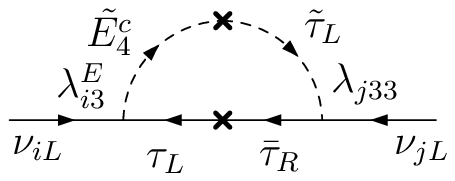}~~~~
\includegraphics[width=2in]{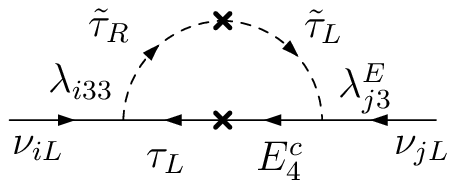}~~~~
\includegraphics[width=2in]{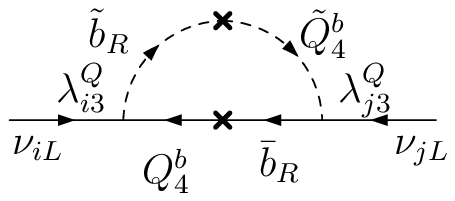}\\
\includegraphics[width=2in]{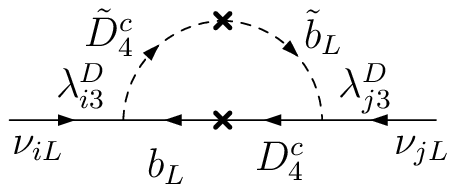}~~~~
\includegraphics[width=2in]{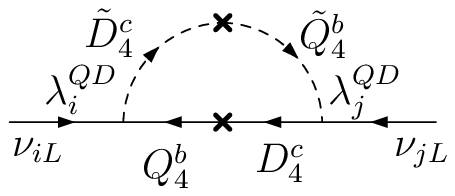}~~~~
\includegraphics[width=2in]{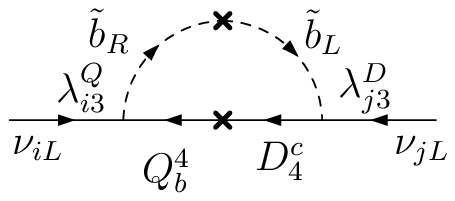}\\
\includegraphics[width=2in]{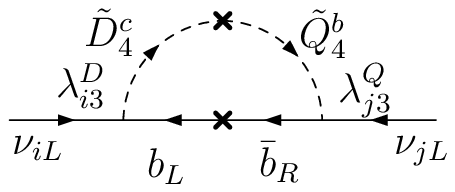}~~~~
\includegraphics[width=2in]{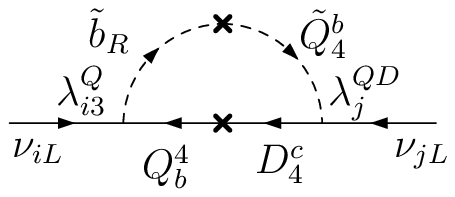}~~~~
\includegraphics[width=2in]{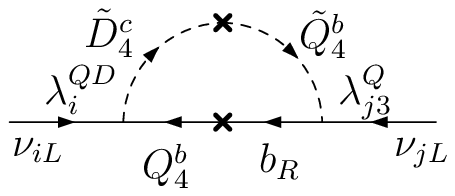}\\
\includegraphics[width=2in]{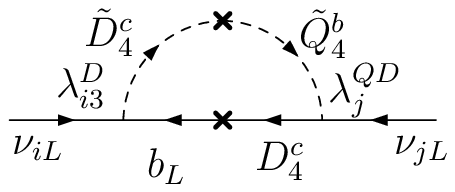}~~~~
\includegraphics[width=2in]{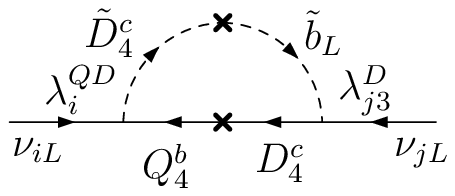}~~~~
\includegraphics[width=2in]{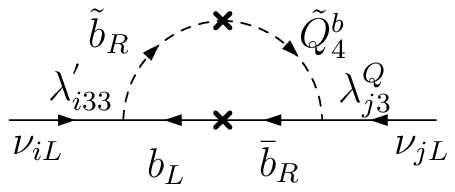}\\
\includegraphics[width=2in]{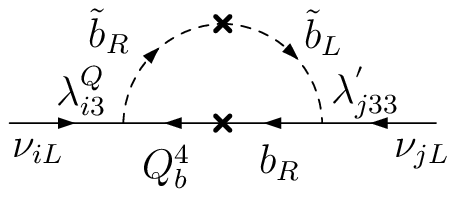}~~~~
\includegraphics[width=2in]{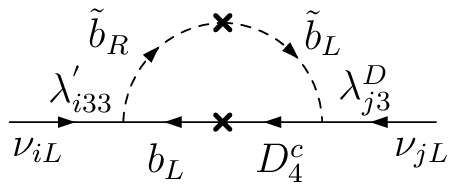}~~~~
\includegraphics[width=2in]{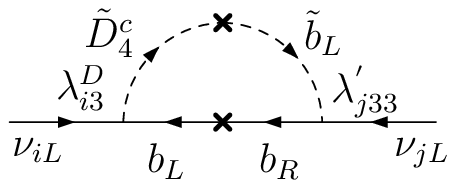}\\
\includegraphics[width=2in]{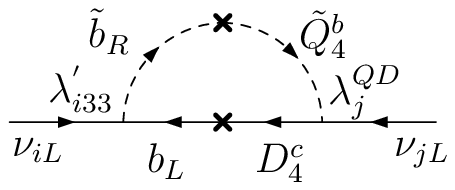}~~~~
\includegraphics[width=2in]{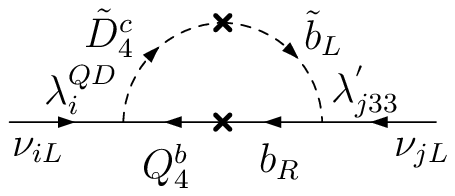}~~~~
\includegraphics[width=2in]{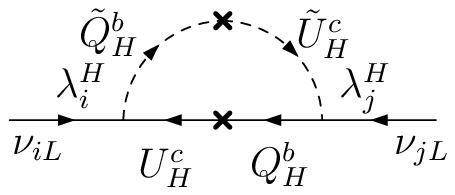}\\
\caption{One-loop contributions to the neutrino masses and mixings in our model.}
\end{figure}

The corresponding analytical results are listed below:
\begin{eqnarray}
 &&M^{\nu}_{ij} |_{\lambda \lambda}\ \simeq\ \frac{1}{8 \pi^2} \sum_{k,m}
  \lambda_{i33} \lambda_{j33}\,R^{*\tau}_{m1} L^{*\tau}_{m1} V^{*\tau}_{k1}
  V^{\tau}_{k2}\ m_{\tau_m}b(m_{\tau_m}, M_{\tilde{\tau}_{L(R)k}}),\\
 && M^{\nu}_{ij} |_{\lambda \lambda^E}\ \simeq\ \frac{1}{8 \pi^2} \sum_{k,m}
  \lambda^E_{i3} \lambda_{j33}\,[R^{*\tau}_{m1} L^{*\tau}_{m1} V^{*\tau}_{k1}
  V^{\tau}_{k3}\ m_{\tau_m}b(m_{\tau_m}, M_{\tilde{\tau}_{L(R)k}})\nonumber\\
 && ~~~~~~~~~~~~~~~~~+R^{*\tau}_{m2} L^{*\tau}_{m1} V^{*\tau}_{k1}
  V^{\tau}_{k2}\ m_{\tau_m}b(m_{\tau_m}, M_{\tilde{\tau}_{L(R)k}})],\\
 &&M^{\nu}_{ij} |_{\lambda^E \lambda^E}\ \simeq\ \frac{1}{8 \pi^2} \sum_{k,m}
  \lambda^E_{i3} \lambda^E_{j3}\,R^{*\tau}_{m2} L^{*\tau}_{m1} V^{*\tau}_{k1}
  V^{\tau}_{k3}\ m_{\tau_m}b(m_{\tau_m}, M_{\tilde{\tau}_{L(R)k}}),\\
&& M^{\nu}_{ij} |_{\lambda' \lambda'}\ \simeq\ \frac{3}{8 \pi^2} \sum_{k,m}
  \lambda'_{i33} \lambda'_{j33}\,[ R^{*b}_{m1} L^{*b}_{m1} V^{*b}_{k1}
  V^{b}_{k2}\ m_{b_m}b(m_{b_m}, M_{\tilde{b}_{L(R)k}})\nonumber\\
  && ~~~~~~~~~~~~~~~~~+\lambda'_{i32} \lambda'_{j23}\,R^{*b}_{m1} L^{*b}_{m1}
  m_{b_m}\sin\alpha_{s1(2)}\cos\alpha_{s1(2)}b(m_{b_m}, M_{\tilde{s}_{1,2}})],\\
&& M^{\nu}_{ij} |_{\lambda^Q \lambda^Q}\ \simeq\ \frac{3}{8 \pi^2} \sum_{k,m}
  \lambda^Q_{i3} \lambda^Q_{j3}\, R^{*b}_{m1} L^{*b}_{m2} V^{*b}_{k2}
  V^{b}_{k5}\ m_{b_m}b(m_{b_m}, M_{\tilde{b}_{L(R)k}}),\\
  && M^{\nu}_{ij} |_{\lambda^D \lambda^D}\ \simeq\ \frac{3}{8 \pi^2} \sum_{k,m}
  \lambda^D_{i3} \lambda^D_{j3}\, R^{*b}_{m2} L^{*b}_{m1} V^{*b}_{k1}
  V^{b}_{k3}\ m_{b_m}b(m_{b_m}, M_{\tilde{b}_{L(R)k}}),\\
  && M^{\nu}_{ij} |_{\lambda^{QD} \lambda^{QD}}\ \simeq\ \frac{3}{8 \pi^2} \sum_{k,m}
  \lambda^{QD}_{i} \lambda^{QD}_{j}\, R^{*b}_{m2} L^{*b}_{m2} V^{*b}_{k3}
  V^{b}_{k5}\ m_{b_m}b(m_{b_m}, M_{\tilde{b}_{L(R)k}}),\\
&& M^{\nu}_{ij} |_{\lambda^Q \lambda^D}\ \simeq\ \frac{3}{8 \pi^2} \sum_{k,m}
  \lambda^Q_{i3} \lambda^D_{j3}\, [R^{*b}_{m2} L^{*b}_{m2} V^{*b}_{k1}
  V^{b}_{k2}\ m_{b_m}b(m_{b_m}, M_{\tilde{b}_{L(R)k}})\nonumber\\
  &&~~~~~~~~~~~~~~~~~+R^{*b}_{m1} L^{*b}_{m1} V^{*b}_{k3}
  V^{b}_{k5}\ m_{b_m}b(m_{b_m}, M_{\tilde{b}_{L(R)k}})],\\
  && M^{\nu}_{ij} |_{\lambda^Q \lambda^{QD}}\ \simeq\ \frac{3}{8 \pi^2} \sum_{k,m}
  \lambda^Q_{i3} \lambda^{QD}_{j}\, [R^{*b}_{m1} L^{*b}_{m2} V^{*b}_{k3}
  V^{b}_{k5}\ m_{b_m}b(m_{b_m}, M_{\tilde{b}_{L(R)k}})\nonumber\\
  &&~~~~~~~~~~~~~~~~~+R^{*b}_{m2} L^{*b}_{m2} V^{*b}_{k2}
  V^{b}_{k5}\ m_{b_m}b(m_{b_m}, M_{\tilde{b}_{L(R)k}})],\\
   && M^{\nu}_{ij} |_{\lambda^D \lambda^{QD}}\ \simeq\ \frac{3}{8 \pi^2} \sum_{k,m}
  \lambda^D_{i3} \lambda^{QD}_{j}\, [R^{*b}_{m2} L^{*b}_{m1} V^{*b}_{k3}
  V^{b}_{k5}\ m_{b_m}b(m_{b_m}, M_{\tilde{b}_{L(R)k}})\nonumber\\
  &&~~~~~~~~~~~~~~~~~+R^{*b}_{m2} L^{*b}_{m2} V^{*b}_{k1}
  V^{b}_{k3}\ m_{b_m}b(m_{b_m}, M_{\tilde{b}_{L(R)k}})],\\
   && M^{\nu}_{ij} |_{\lambda^{'} \lambda^Q}\ \simeq\ \frac{3}{8 \pi^2} \sum_{k,m}
  \lambda^{'}_{i33} \lambda^Q_{j3}\, [R^{*b}_{m1} L^{*b}_{m1} V^{*b}_{k2}
  V^{b}_{k5}\ m_{b_m}b(m_{b_m}, M_{\tilde{b}_{L(R)k}})\nonumber\\
  &&~~~~~~~~~~~~~~~~~+R^{*b}_{m1} L^{*b}_{m2} V^{*b}_{k1}
  V^{b}_{k2}\ m_{b_m}b(m_{b_m}, M_{\tilde{b}_{L(R)k}})],\\
    && M^{\nu}_{ij} |_{\lambda^{'} \lambda^D}\ \simeq\ \frac{3}{8 \pi^2} \sum_{k,m}
  \lambda^{'}_{i33} \lambda^D_{j3}\, [R^{*b}_{m2} L^{*b}_{m1} V^{*b}_{k1}
  V^{b}_{k2}\ m_{b_m}b(m_{b_m}, M_{\tilde{b}_{L(R)k}})\nonumber\\
  &&~~~~~~~~~~~~~~~~~+R^{*b}_{m1} L^{*b}_{m1} V^{*b}_{k1}
  V^{b}_{k3}\ m_{b_m}b(m_{b_m}, M_{\tilde{b}_{L(R)k}})],
\end{eqnarray}
\begin{eqnarray}
    && M^{\nu}_{ij} |_{\lambda^{'} \lambda^{QD}}\ \simeq\ \frac{3}{8 \pi^2} \sum_{k,m}
  \lambda^{'}_{i33} \lambda^{QD}_{j}\, [R^{*b}_{m2} L^{*b}_{m1} V^{*b}_{k2}
  V^{b}_{k5}\ m_{b_m}b(m_{b_m}, M_{\tilde{b}_{L(R)k}})\nonumber\\
  &&~~~~~~~~~~~~~~~~~+R^{*b}_{m1} L^{*b}_{m2} V^{*b}_{k1}
  V^{b}_{k3}\ m_{b_m}b(m_{b_m}, M_{\tilde{b}_{L(R)k}})],\\
 &&M^{\nu}_{ij} |_{\lambda^H \lambda^H}\ \simeq\ \frac{3}{8 \pi^2} \sum_{k,m}
  \lambda^H_{i} \lambda^H_{j}\, R^{*t}_{m3} L^{*t}_{m3} V^{*u}_{k6}
  V^{u}_{k4}\ m_{t_m}b(m_{t_m}, M_{\tilde{t}_{L(R)k}}),
\end{eqnarray}
In which $L^{\tau, b, t},R^{\tau, b, t}$ are biunitary
matrices of mass matrices between ($\tau, b, t$) and the
vector-like fermions (see Appendix A), while
$m_{\tau_m}, m_{b_m},m_{t_m}$
indicate the corresponding mass eigenvalues.
$V^{\tilde{\tau}, \tilde{d}, \tilde{t}}$ are the square mass
mixing unitary matrices of their superpatners, while
$M_{\tilde{\tau}_{L(R)k}}, M_{\tilde{b}_{L(R)k}},
M_{\tilde{t}_{L(R)k}}$ stand for the corresponding mass
eigenvalues.  $\sin\alpha_{s1(2)}$, $\cos\alpha_{s1(2)}$ are the
unitary matrix elements of $\tilde{s}$. $b(m_1, m_2)$ is the loop
integral factor: $b(m_1, m_2)\equiv\frac{1}{m^2_1-m^2_2}(m^2_1\ln
m^2_1-m^2_2\ln m^2_2-m^2_2+m^2_1 )$. The value range of the
indices in Eq. (4)-(6) is m=1,2, k=1-4 ,while in Eq. (7)-(17), it
is m=1,2,3, k=1-6.

\section{NEUTRINO SPECTRUM-CALCULATING METHOD AND PARAMETER SETTINGS}

The methods to generate neutrino masses and mixing angles with
one-loop trilinear $\slashed L$ couplings actually involves the
following three matrices
\begin{eqnarray}
m_1
\begin{pmatrix}
a^2~~ ab~~ ac \\ ab~~ b^2~~ bc \\ ac~~ bc~~ c^2
\end{pmatrix}
~,~
m_2
\begin{pmatrix}
d^2~~ de~~ df \\ de~~ e^2~~ ef \\ df~~ ef~~ f^2
\end{pmatrix}
~,~
m_3
\begin{pmatrix}
g^2~~ gh~~ gl \\ gh~~ h^2~~ hl \\ gl~~ hl~~ l^2
\end{pmatrix}
~,~
\end{eqnarray}
where we name each of the matrices above $\mathcal{M}_{1,2,3}$
separately. We assume $m_1> m_{2,3}$, $m_2\sim m_3$ and there is
no strong hierarchy between a, b, c, d, e, f, g, h, l.

$\mathcal{M}_1$ has only one eigenvalue after digonalized by an
unitary rotation
\begin{equation}
X^{T}\mathcal{M}_1 X=\mathrm{diag}(0,~0,~M_1)~,
\end{equation}
where
\begin{equation}
M_1=m_1 (a^2+b^2+c^2)~,
\end{equation}
and
\begin{eqnarray}
X=
\begin{pmatrix}
c_2~~ s_2c_3~~ s_2s_3 \\ -s_2~~ c_2c_3~~ s_2s_3 \\ 0~~ -s_3~~ c_3
\end{pmatrix}
~,~
\end{eqnarray}
\begin{eqnarray}
s_2=\frac{a}{\sqrt{a^2+b^2}}~,~c_3=\frac{c}{\sqrt{a^2+b^2+c^2}}.
\end{eqnarray}

If we rotate the sum over $\mathcal{M}_{1,2,3}$ by matrix X, it
becomes
\begin{equation}
X^{T}(\mathcal{M}_1+\mathcal{M}_2+\mathcal{M}_3) X\approx
m_1 (a^2+b^2+c^2)
\begin{pmatrix}
\epsilon_{11}~~ \epsilon_{12}~~ \epsilon_{13} \\
\epsilon_{21}~~\epsilon_{22}~~ \epsilon_{23} \\
\epsilon_{31}~~ \epsilon_{32}~~ 1
\end{pmatrix}
~,~
\end{equation}
where $\epsilon_{ij}$ are some small values related with
$m_2/m_1,m_3/m_1$ and the other elements of $\mathcal{M}_{1,2,3}$.
We can then define another unitary matrix $X'$ to diagonalize
the matrix in Eq. (B6) in an approximate way:
\begin{eqnarray}
X^{'T}X^{T}(\mathcal{M}_1+\mathcal{M}_2+\mathcal{M}_3) XX^{'}
&\approx m_1 (a^2+b^2+c^2)~
\mathrm{diag}(\delta'_3,~\delta'_2,~1)
~,~
\end{eqnarray}
where
\begin{equation}
X'=
\begin{pmatrix}
c_1& s_1&0 \\
-s_1&c_1& 0 \\
0& 0& 1
\end{pmatrix}
~,~
\end{equation}
and
\begin{equation}
\tan2\theta_1=\frac{2\epsilon_{12}}{\epsilon_{22}-\epsilon_{11}}~.~
\end{equation}
Then from Eq. (B7), we get all three mass eigenvalues
\begin{equation}
M_1\sim m_1 (a^2+b^2+c^2)~,~M_2\sim M_1 \delta'_2~,~M_3\sim M_1 \delta'_3~,~
\end{equation}
and from Eq. (B4, B8), we get all three mixing angles
\begin{eqnarray}
&s_{13}=s_2s_3=\frac{a}{\sqrt{a^2+b^2+c^2}}~,~\nonumber\\
&s_{23}=c_2s_3/c_{13}=\frac{b}{\sqrt{b^2+c^2}}~,~\\
&s_{12}=(s_1c_2+c_1s_2c_3)/c_{13}~.~\nonumber
\end{eqnarray}

The parameter settings we used in table I are given as following

\centerline{Set I:}
\begin{eqnarray}
&&m^{\tau}_{34} =10,~M_E = M_{EH}=600\mathrm{GeV},~B^E \mu^E= 400^2\mathrm{GeV}^2
;~m^b_H=170\mathrm{GeV}, \nonumber \\
&&m^{b}_{34} = m^{b}_{43}=m^{b}_{44}=0,~
M_Q = M_D=M_{DH}=M_{QH}=700\mathrm{GeV},\nonumber \\
&&B^D \mu^D= B^Q \mu^Q=500^2\mathrm{GeV}^2;~
m^{t}_{34} = m^{t}_{43}=m^t_H=13\mathrm{GeV},~ m^{t}_{44}=174\mathrm{GeV},\nonumber \\
&&M_U=M_{UH}=700\mathrm{GeV},~
B^U \mu^U=500^2\mathrm{GeV}^2,~
\tan \beta= 10,~A=\mu=500\mathrm{GeV}.\nonumber
\end{eqnarray}

\centerline{Set II:}
\begin{eqnarray}
&&m^{\tau}_{34} =10,~M_E = M_{EH}=1000\mathrm{GeV},~B^E \mu^E= 600^2\mathrm{GeV}^2
;~m^b_H=170\mathrm{GeV}, \nonumber \\
&&m^{b}_{34} = m^{b}_{43}=m^{b}_{44}=10\mathrm{GeV},~
M_Q = M_D=M_{DH}=M_{QH}=1000\mathrm{GeV},\nonumber \\
&&B^D \mu^D= B^Q \mu^Q=600^2\mathrm{GeV}^2;~
m^{t}_{34} = m^{t}_{43}=m^t_H=13\mathrm{GeV},~ m^{t}_{44}=174\mathrm{GeV},\nonumber \\
&&M_U=M_{UH}=1000\mathrm{GeV},~
B^U \mu^U=600^2\mathrm{GeV}^2,~
\tan \beta= 10,~A=\mu=600\mathrm{GeV}.\nonumber
\end{eqnarray}

\section{EXOTIC QUARK AND LEPTON COUPLINGS TO $W, Z, h^0$ AND DECAY WIDTHS}

The couplings for the W, Z, $h^0$ boson with leptons in Eq. (12) are
\begin{eqnarray}
&&g^W_{\bar{\tau}_{1L}\nu_{\tau L}}=\frac{g}{\sqrt{2}} L^\tau_{21}~~,~~\nonumber\\
&&g^Z_{\bar{\tau}_{1L}\tau_{L}}=\frac{g s^2_{W}}{c_{W}}L^\tau_{22}L^\tau_{12}
-\frac{g}{4c_{W}}[(2-4s^2_{W})L^\tau_{21}L^\tau_{11}]~~,\nonumber\\
&&g^Z_{\bar{\tau}_{1R}\tau_{R}}=\frac{g s^2_{W}}{c_{W}}R^\tau_{22}R^\tau_{12}
+\frac{g}{4c_{W}}(4s^2_{W}R^\tau_{21}R^\tau_{11})~~,\\
&&g^{h^0}_{\bar{\tau}_{1L}\tau_{R}}=-\frac{s_\alpha}{\sqrt{2}}(y^\tau_{33}L^\tau_{21}R^\tau_{11}
+y^E_{3}L^\tau_{21}R^\tau_{12})~~,~~\nonumber\\
&&g^{h^0}_{\bar{\tau}_{L}\tau_{1R}}=-\frac{s_\alpha}{\sqrt{2}}(y^\tau_{33}L^\tau_{11}R^\tau_{21}
+y^E_{3}L^\tau_{11}R^\tau_{22})~~,\nonumber
\end{eqnarray}
where $c_\alpha=s_\beta$, $s_\alpha=-c_\beta$ is the elements of the rotation
matrix related with the real parts of $(H^0_u,~H^0_d)$.

Then the decay widths of $\tau_{1}$ are
\begin{eqnarray}
&&\Gamma(\mathrm{\tau_{1}\rightarrow W \nu_\tau})=\frac{m_{\tau_{1}}}{32\pi}
(1+x_W^4-2x_W^2)^{1/2}(1-2x_W^2+x_W^{-2})(g^{W}_{\bar{\tau}_{1}\nu_{\tau }})^2~~,~~\nonumber\\
&&\Gamma(\mathrm{\tau_{1}\rightarrow Z \tau_{}})=\frac{m_{\tau_{1}}}{32\pi}
(1+x_Z^4+x_{\tau_{}}^4-2x_Z^2-2x_{\tau_{}}^2-2x_Z^2x_{\tau_{}}^2)^{1/2}\nonumber\\
&&~~~~~~~~~~~~~~~~~~~~
\{(1+x_{\tau_{}}^2-2x_Z^2+(1-x_{\tau_{}}^2)^2x_Z^{-2})[(g^{Z}_{\bar{\tau}_{1L}\tau_{L}})^2
+(g^{Z}_{\bar{\tau}_{1R}\tau_{R}})^2]\nonumber\\
&&~~~~~~~~~~~~~~~~~~~~
+12x_{\tau_{}}g^{Z}_{\bar{\tau}_{1L}\tau_{L}}g^{Z}_{\bar{\tau}_{1R}\tau_{R}}
\}~~,~~\\
&&\Gamma(\mathrm{\tau_{1}\rightarrow h^0 \tau_{}})=\frac{m_{\tau_{1}}}{32\pi}
(1+x_{h^0}^4+x_{\tau_{}}^4-2x_{h^0}^2-2x_{\tau_{}}^2-2x_{h^0}^2x_{\tau_{}}^2)^{1/2}\nonumber\\
&&~~~~~~~~~~~~~~~~~~~~
\{(1+x_{\tau_{}}^2-x_{h^0}^2)[(g^{h^0}_{\bar{\tau}_{1L}\tau_{R}})^2
+(g^{h^0}_{\bar{\tau}_{L}\tau_{1R}})^2]\nonumber\\
&&~~~~~~~~~~~~~~~~~~~~
+4x_{\tau_{}}g^{h^0}_{\bar{\tau}_{1L}\tau_{R}}g^{h^0}_{\bar{\tau}_{L}\tau_{1R}}
\}~~,~~\nonumber
\end{eqnarray}
where $x_i=m_i/m_{\tau_{1}}$ for $i=\mathrm{W, Z, \tau_{}, h^0}$.

The couplings for the W, Z, $h^0$ boson with $t, t_1, t_2$ in Eq. (13) are
\begin{eqnarray}
&&g^W_{\bar{t}_{1L}b_{L}}=\frac{g}{\sqrt{2}} (L^t_{21}L^b_{11}+L^t_{22}L^b_{12})~~,~~
g^W_{\bar{t}_{1R}b_{R}}=\frac{g}{\sqrt{2}} R^t_{23}R^b_{13}~~,\nonumber\\
&&g^W_{\bar{t}_{2L}b_{L}}=\frac{g}{\sqrt{2}} (L^t_{31}L^b_{11}+L^t_{32}L^b_{12})~~,~~
g^W_{\bar{t}_{2R}b_{R}}=\frac{g}{\sqrt{2}} R^t_{33}R^b_{13}~~,\nonumber\\
&&g^W_{\bar{t}_{2L}b_{1L}}=\frac{g}{\sqrt{2}} (L^t_{31}L^b_{21}+L^t_{32}L^b_{22})~~,~~
g^W_{\bar{t}_{2R}b_{1R}}=\frac{g}{\sqrt{2}} R^t_{33}R^b_{23}~~,\nonumber\\
&&g^Z_{\bar{t}_{1L}t_{L}}=\frac{-2g s^2_{W}}{3c_{W}}L^t_{23}L^t_{13}
+\frac{g}{4c_{W}}[(2-\frac{8}{3}s^2_{W})(L^t_{21}L^t_{11}+L^t_{22}L^t_{12})]~~,\nonumber\\
&&g^Z_{\bar{t}_{1R}t_{R}}=
-\frac{g}{4c_{W}}[\frac{8}{3}s^2_{W}(R^t_{21}R^t_{11}+R^t_{22}R^t_{12})+
(2-\frac{4}{3}s^2_{W})R^t_{23}R^t_{13}]~~,\nonumber\\
&&g^Z_{\bar{t}_{2L}t_{L}}=\frac{-2g s^2_{W}}{3c_{W}}L^t_{33}L^t_{13}
+\frac{g}{4c_{W}}[(2-\frac{8}{3}s^2_{W})(L^t_{31}L^t_{11}+L^t_{32}L^t_{12})]~~,\\
&&g^Z_{\bar{t}_{2R}t_{R}}=
-\frac{g}{4c_{W}}[\frac{8}{3}s^2_{W}(R^t_{31}R^t_{11}+R^t_{32}R^t_{12})+
(2-\frac{4}{3}s^2_{W})R^t_{33}R^t_{13}]~~,\nonumber\\
&&g^Z_{\bar{t}_{2L}t_{1L}}=\frac{-2g s^2_{W}}{3c_{W}}L^t_{33}L^t_{23}
+\frac{g}{4c_{W}}[(2-\frac{8}{3}s^2_{W})(L^t_{31}L^t_{21}+L^t_{32}L^t_{22})]~~,\nonumber\\
&&g^Z_{\bar{t}_{2R}t_{1R}}=
-\frac{g}{4c_{W}}[\frac{8}{3}s^2_{W}(R^t_{31}R^t_{21}+R^t_{32}R^t_{22})+
(2-\frac{4}{3}s^2_{W})R^t_{33}R^t_{23}]~~,\nonumber\\
&&g^{h^0}_{\bar{t}_{1L}t_{R}}=\frac{c_\alpha}{\sqrt{2}}(y^u_{33}L^t_{21}R^t_{11}
+y^{Q}_{3}L^t_{22}R^t_{11}+y^U_{3}L^t_{21}R^t_{12}+y^{QU}_{3}L^t_{22}R^t_{12})
-\frac{s_\alpha}{\sqrt{2}}y^{H}L^t_{23}R^t_{13}~~,\nonumber\\
&&g^{h^0}_{\bar{t}_{L}t_{1R}}=\frac{c_\alpha}{\sqrt{2}}(y^u_{33}L^t_{11}R^t_{21}
+y^{Q}_{3}L^t_{12}R^t_{21}+y^U_{3}L^t_{11}R^t_{22}+y^{QU}_{3}L^t_{12}R^t_{22})
-\frac{s_\alpha}{\sqrt{2}}y^{H}L^t_{13}R^t_{23}~~,\nonumber\\
&&g^{h^0}_{\bar{t}_{2L}t_{R}}=\frac{c_\alpha}{\sqrt{2}}(y^u_{33}L^t_{31}R^t_{11}
+y^{Q}_{3}L^t_{32}R^t_{11}+y^U_{3}L^t_{31}R^t_{12}+y^{QU}_{3}L^t_{32}R^t_{12})
-\frac{s_\alpha}{\sqrt{2}}y^{H}L^t_{33}R^t_{13}~~,\nonumber\\
&&g^{h^0}_{\bar{t}_{L}t_{2R}}=\frac{c_\alpha}{\sqrt{2}}(y^u_{33}L^t_{11}R^t_{31}
+y^{Q}_{3}L^t_{12}R^t_{31}+y^U_{3}L^t_{11}R^t_{32}+y^{QU}_{3}L^t_{12}R^t_{32})
-\frac{s_\alpha}{\sqrt{2}}y^{H}L^t_{13}R^t_{33}~~,\nonumber\\
&&g^{h^0}_{\bar{t}_{2L}t_{1R}}=\frac{c_\alpha}{\sqrt{2}}(y^u_{33}L^t_{31}R^t_{21}
+y^{Q}_{3}L^t_{32}R^t_{21}+y^U_{3}L^t_{31}R^t_{22}+y^{QU}_{3}L^t_{32}R^t_{22})
-\frac{s_\alpha}{\sqrt{2}}y^{H}L^t_{33}R^t_{23}~~,\nonumber\\
&&g^{h^0}_{\bar{t}_{1L}t_{2R}}=\frac{c_\alpha}{\sqrt{2}}(y^u_{33}L^t_{21}R^t_{31}
+y^{Q}_{3}L^t_{22}R^t_{31}+y^U_{3}L^t_{21}R^t_{32}+y^{QU}_{3}L^t_{22}R^t_{32})
-\frac{s_\alpha}{\sqrt{2}}y^{H}L^t_{23}R^t_{33}~~.\nonumber
\end{eqnarray}

The decay widths of the lightest new up-type quark
$\mathrm{t_{1}}$ are
\begin{eqnarray}
&&\Gamma(\mathrm{t_{1}\rightarrow W b_{}})=\frac{m_{t_{1}}}{32\pi}
(1+x_W^4+x_{b_{}}^4-2x_W^2-2x_{b_{}}^2-2x_W^2x_{b_{}}^2)^{1/2}\nonumber\\
&&~~~~~~~~~~~~~~~~~~~~
\{(1+x_{b_{}}^2-2x_W^2+(1-x_{b_{}}^2)^2x_W^{-2})[(g^{W}_{\bar{t}_{1L}b_{L}})^2
+(g^{W}_{\bar{t}_{1R}b_{R}})^2]\nonumber\\
&&~~~~~~~~~~~~~~~~~~~~
+12x_{b_{}}g^{W}_{\bar{t}_{1L}b_{L}}g^{W}_{\bar{t}_{1R}b_{R}}
\}~~,~~\nonumber\\
&&\Gamma(\mathrm{t_{1}\rightarrow Z t_{}})=\frac{m_{t_{1}}}{32\pi}
(1+x_Z^4+x_{t_{}}^4-2x_Z^2-2x_{t_{}}^2-2x_Z^2x_{t_{}}^2)^{1/2}\nonumber\\
&&~~~~~~~~~~~~~~~~~~~~
\{(1+x_{t_{}}^2-2x_Z^2+(1-x_{t_{}}^2)^2x_Z^{-2})[(g^{'Z}_{\bar{t}_{1L}t_{L}})^2
+(g^{Z}_{\bar{t}_{1R}t_{R}})^2]\\
&&~~~~~~~~~~~~~~~~~~~~
+12x_{t_{}}g^{Z}_{\bar{t}_{1L}t_{L}}g^{Z}_{\bar{t}_{1R}t_{R}}
\}~~,~~\nonumber
\end{eqnarray}
\begin{eqnarray}
&&\Gamma(\mathrm{t_{1}\rightarrow h^0 t_{}})=\frac{m_{t_{1}}}{32\pi}
(1+x_{h^0}^4+x_{t_{}}^4-2x_{h^0}^2-2x_{t_{}}^2-2x_{h^0}^2x_{t_{}}^2)^{1/2}\nonumber\\
&&~~~~~~~~~~~~~~~~~~~~
\{(1+x_{t_{}}^2-x_{h^0}^2)[(g^{h^0}_{\bar{t}_{1L}t_{R}})^2
+(g^{h^0}_{\bar{t}_{L}t_{1R}})^2]\nonumber\\
&&~~~~~~~~~~~~~~~~~~~~
+4x_{t_{}}g^{h^0}_{\bar{t}_{1L}t_{R}}g^{h^0}_{\bar{t}_{L}t_{1R}}
\}~~,~~\nonumber
\end{eqnarray}
where $x_i=m_i/m_{t_{1}}$ for $i=\mathrm{W, Z, t_{}, h^0}$. The
heaviest new up-type quark $\mathrm{t_{2}}$~ has six decay
channels. The decay widths have the similar forms and can be
deduced straightforwardly.

The couplings for the W, Z, $h^0$ boson with $b, b_1, b_2$ in Eq. (14) are
\begin{eqnarray}
&&g^W_{\bar{b}_{1L}t_{L}}=\frac{g}{\sqrt{2}} (L^b_{21}L^t_{11}+L^b_{22}L^t_{12})~~,~~
g^W_{\bar{b}_{1R}t_{R}}=\frac{g}{\sqrt{2}} R^b_{23}R^t_{13}~~,\nonumber\\
&&g^W_{\bar{b}_{2L}t_{L}}=\frac{g}{\sqrt{2}} (L^b_{31}L^t_{11}+L^b_{32}L^t_{12})~~,~~
g^W_{\bar{b}_{2R}t_{R}}=\frac{g}{\sqrt{2}} R^b_{33}R^t_{13}~~,\nonumber\\
&&g^W_{\bar{b}_{2L}t_{1L}}=\frac{g}{\sqrt{2}} (L^b_{31}L^t_{21}+L^b_{32}L^t_{22})~~,~~
g^W_{\bar{b}_{2R}t_{1R}}=\frac{g}{\sqrt{2}} R^b_{33}R^t_{23}~~,\nonumber\\
&&g^Z_{\bar{b}_{1L}b_{L}}=\frac{g s^2_{W}}{3c_{W}}L^b_{23}L^b_{13}
-\frac{g}{4c_{W}}[(2-\frac{4}{3}s^2_{W})(L^b_{21}L^b_{11}+L^b_{22}L^b_{12})]~~,\nonumber\\
&&g^Z_{\bar{b}_{1R}b_{R}}=\frac{g}{4c_{W}}[\frac{4}{3}s^2_{W}(R^b_{21}R^b_{11}+R^b_{22}R^b_{12})
+(2-\frac{8}{3}s^2_{W})R^b_{23}R^b_{13}]~~,\nonumber\\
&&g^Z_{\bar{b}_{2L}b_{L}}=\frac{g s^2_{W}}{3c_{W}}L^b_{33}L^b_{13}
-\frac{g}{4c_{W}}[(2-\frac{4}{3}s^2_{W})(L^b_{31}L^b_{11}+L^b_{32}L^b_{12})]~~,\nonumber\\
&&g^Z_{\bar{b}_{2R}b_{R}}=\frac{g}{4c_{W}}[\frac{4}{3}s^2_{W}(R^b_{31}R^b_{11}+R^b_{32}R^b_{12})+
(2-\frac{8}{3}s^2_{W})R^b_{33}R^b_{13}]~~,\nonumber\\
&&g^Z_{\bar{b}_{2L}b_{1L}}=\frac{g s^2_{W}}{3c_{W}}L^b_{33}L^b_{23}
-\frac{g}{4c_{W}}[(2-\frac{4}{3}s^2_{W})(L^b_{31}L^b_{21}+L^b_{32}L^b_{22})]~~,\\
&&g^Z_{\bar{b}_{2R}b_{1R}}=\frac{g}{4c_{W}}[\frac{4}{3}s^2_{W}(R^b_{31}R^b_{21}+R^b_{32}R^b_{22})+
(2-\frac{8}{3}s^2_{W})R^b_{33}R^b_{23}]~~,\nonumber\\
&&g^{h^0}_{\bar{b}_{1L}b_{R}}=-\frac{s_\alpha}{\sqrt{2}}(y^d_{33}L^b_{21}R^b_{11}
+y^{Q'}_{3}L^b_{22}R^b_{11}+y^D_{3}L^b_{21}R^b_{12}+y^{QD}_{3}L^b_{22}R^b_{12})
+\frac{c_\alpha}{\sqrt{2}}y^{H'}L^b_{23}R^b_{13}~~,\nonumber\\
&&g^{h^0}_{\bar{b}_{L}b_{1R}}=-\frac{s_\alpha}{\sqrt{2}}(y^d_{33}L^b_{11}R^b_{21}
+y^{Q'}_{3}L^b_{12}R^b_{21}+y^D_{3}L^b_{11}R^b_{22}+y^{QD}_{3}L^b_{12}R^b_{22})
+\frac{c_\alpha}{\sqrt{2}}y^{H'}L^b_{13}R^b_{23}~~,\nonumber\\
&&g^{h^0}_{\bar{b}_{2L}b_{R}}=-\frac{s_\alpha}{\sqrt{2}}(y^d_{33}L^b_{31}R^b_{11}
+y^{Q'}_{3}L^b_{32}R^b_{11}+y^D_{3}L^b_{31}R^b_{12}+y^{QD}_{3}L^b_{32}R^b_{12})
+\frac{c_\alpha}{\sqrt{2}}y^{H'}L^b_{33}R^b_{13}~~,\nonumber\\
&&g^{h^0}_{\bar{b}_{L}b_{2R}}=-\frac{s_\alpha}{\sqrt{2}}(y^d_{33}L^b_{11}R^b_{31}
+y^{Q'}_{3}L^b_{12}R^b_{31}+y^D_{3}L^b_{11}R^b_{32}+y^{QD}_{3}L^b_{12}R^b_{32})
+\frac{c_\alpha}{\sqrt{2}}y^{H'}L^b_{13}R^b_{33}~~,\nonumber\\
&&g^{h^0}_{\bar{b}_{2L}b_{1R}}=-\frac{s_\alpha}{\sqrt{2}}(y^d_{33}L^b_{31}R^b_{21}
+y^{Q'}_{3}L^b_{32}R^b_{21}+y^D_{3}L^b_{31}R^b_{22}+y^{QD}_{3}L^b_{32}R^b_{22})
+\frac{c_\alpha}{\sqrt{2}}y^{H'}L^b_{33}R^b_{23}~~,\nonumber\\
&&g^{h^0}_{\bar{b}_{1L}b_{2R}}=-\frac{s_\alpha}{\sqrt{2}}(y^d_{33}L^b_{21}R^b_{31}
+y^{Q'}_{3}L^b_{22}R^b_{31}+y^D_{3}L^b_{21}R^b_{32}+y^{QD}_{3}L^b_{22}R^b_{32})
+\frac{c_\alpha}{\sqrt{2}}y^{H'}L^b_{23}R^b_{33}~~.\nonumber
\end{eqnarray}

The decay widths of the lightest new down-type quark
$\mathrm{b_{1}}$ are
\begin{eqnarray}
&&\Gamma(\mathrm{b_{1}\rightarrow W t})=\frac{m_{b_{1}}}{32\pi}
(1+x_W^4+x_{t_{}}^4-2x_W^2-2x_{t_{}}^2-2x_W^2x_{t_{}}^2)^{1/2}\nonumber\\
&&~~~~~~~~~~~~~~~~~~~~
\{(1+x_{t_{}}^2-2x_W^2+(1-x_{t_{}}^2)^2x_W^{-2})[(g^{W}_{\bar{b}_{1L}t_{L}})^2
+(g^{W}_{\bar{b}_{1R}t_{R}})^2]\nonumber\\
&&~~~~~~~~~~~~~~~~~~~~
+12x_{t_{}}g^{W}_{\bar{b}_{1L}t_{L}}g^{W}_{\bar{b}_{1R}t_{R}}
\}~~,~~\nonumber\\
&&\Gamma(\mathrm{b_{1}\rightarrow Z b})=\frac{m_{b_{1}}}{32\pi}
(1+x_Z^4+x_{b_{}}^4-2x_Z^2-2x_{b_{}}^2-2x_Z^2x_{b_{}}^2)^{1/2}\nonumber\\
&&~~~~~~~~~~~~~~~~~~~~
\{(1+x_{b_{}}^2-2x_Z^2+(1-x_{b_{}}^2)^2x_Z^{-2})[(g^{Z}_{\bar{t}_{1L}t_{L}})^2
+(g^{Z}_{\bar{b}_{1R}b_{R}})^2]\\
&&~~~~~~~~~~~~~~~~~~~~
+12x_{b_{}}g^{Z}_{\bar{b}_{1L}b_{L}}g^{Z}_{\bar{b}_{1R}b_{R}}
\}~~,~~\nonumber\\
&&\Gamma(\mathrm{b_{1}\rightarrow h^0 b})=\frac{m_{b_{1}}}{32\pi}
(1+x_{h^0}^4+x_{b_{}}^4-2x_{h^0}^2-2x_{b_{}}^2-2x_{h^0}^2x_{b_{}}^2)^{1/2}\nonumber\\
&&~~~~~~~~~~~~~~~~~~~~
\{(1+x_{b_{}}^2-x_{h^0}^2)[(g^{h^0}_{\bar{b}_{1L}b_{R}})^2
+(g^{h^0}_{\bar{b}_{L}b_{1R}})^2]\nonumber\\
&&~~~~~~~~~~~~~~~~~~~~
+4x_{b_{}}g^{h^0}_{\bar{b}_{1L}b_{R}}g^{h^0}_{\bar{b}_{L}b_{1R}}
\}~~,~~\nonumber
\end{eqnarray}
where $x_i=m_i/m_{b_{1}}$ for index $i=\mathrm{W, Z, b, h^0}$.
The heaviest new down-type quark~$\mathrm{b_{2}}$~ has six decay
channels, the decay widths have the similar forms and can be
deduced straightforwardly.

\end{document}